\documentclass[journal]{IEEEtran}
\usepackage{cite}
\usepackage{amsmath,amssymb,amsfonts}
\usepackage{algorithmic}
\usepackage{graphicx,color}
\usepackage{textcomp}
\usepackage{xcolor}
\usepackage{hyperref}
\hypersetup{hidelinks=true}
\usepackage{algorithm,algorithmic}
\def\BibTeX{{\rm B\kern-.05em{\sc i\kern-.025em b}\kern-.08em
    T\kern-.1667em\lower.7ex\hbox{E}\kern-.125emX}}
\AtBeginDocument{\definecolor{tmlcncolor}{cmyk}{0.93,0.59,0.15,0.02}\definecolor{NavyBlue}{RGB}{0,86,125}}

\usepackage{cite}
\usepackage{bookmark}
\usepackage{subfig}
\usepackage{booktabs}

\usepackage{bbm}
\usepackage{textcomp}
\usepackage{mathtools}
\usepackage{stfloats}
\usepackage{cuted}
\usepackage{bm}

\usepackage{flushend}
\flushend

\def\OJlogo{\vspace{-4pt}\includegraphics[height=18pt]{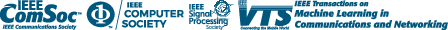}}
\def\seclogo{\vspace{10pt}\includegraphics[height=18pt]{new_logo}}

\DeclarePairedDelimiter\abs{\lvert}{\rvert}

\begin{document}


\title{Full-Duplex Millimeter Wave MIMO Channel Estimation: A Neural Network Approach}

\author{Mehdi~Sattari,~\IEEEmembership{Graduate Student Member,~IEEE,}
        Hao~Guo,~\IEEEmembership{Member,~IEEE,}
        Deniz~Gündüz,~\IEEEmembership{Fellow,~IEEE,}
        Ashkan~Panahi,~\IEEEmembership{Senior Member,~IEEE,}
        and
        Tommy~Svensson,~\IEEEmembership{Senior Member,~IEEE}
        ~
\thanks{This work has been supported, in part, by the project SEMANTIC, funded by the EU’s Horizon 2020 research and innovation programme under the Marie Skłodowska-Curie grant agreement No 861165, and by the Swedish Research Council (VR) grant 2023-00272. The computations were enabled by resources provided by the National Academic Infrastructure for Supercomputing in Sweden (NAISS), partially funded by the VR through grant agreement no. 2022-06725. Corresponding author: M. Sattari. }
\thanks{M. Sattari, H. Guo, and T. Svensson are with the Department
of Electrical Engineering, Chalmers University of Technology, Gothenburg,
Sweden. H.~Guo is also with Electrical and Computer Engineering Department, New York University Tandon School of Engineering, Brooklyn, NY 11201, USA. E-mail: \{mehdi.sattari, hao.guo, tommy.svensson\}@chalmers.se.}
\thanks{A. Panahi is with the Department
of Computer Science, Chalmers University of Technology, Gothenburg,
Sweden. E-mail: ashkan.panahi@chalmers.se.}
\thanks{D. Gunduz is with the Department
of Electrical and Electronic Engineering, Imperial College London, London,
UK. E-mail: d.gunduz@imperial.ac.uk.}}

\maketitle
\begin{abstract}
Millimeter wave (mmWave) multiple-input-multi-output (MIMO) is now a reality
with great potential for further improvement. 
We study full-duplex transmissions as an effective way to improve mmWave MIMO systems. Compared to half-duplex systems, full-duplex transmissions may offer higher data rates and lower latency.
However, full-duplex transmission is hindered by self-interference (SI) 
at the receive antennas, and SI channel estimation becomes a crucial step to make the full-duplex systems feasible.
In this paper, we address the problem of channel estimation in full-duplex mmWave MIMO systems using neural networks (NNs). Our approach involves sharing pilot resources between user equipments (UEs) and transmit antennas at the base station (BS), aiming to reduce the pilot overhead in full-duplex systems and to achieve a comparable level to that of a half-duplex system. Additionally, in the case of separate antenna configurations in a full-duplex BS, providing channel estimates of transmit antenna (TX) arrays to the downlink UEs poses another challenge, as the TX arrays are not capable of receiving pilot signals. To address this, we employ an NN to map the channel from the downlink UEs to the receive antenna (RX) arrays to the channel from the TX arrays to the downlink UEs. 
We further elaborate on how NNs perform the estimation with different architectures, (e.g., different numbers of hidden layers), the introduction of non-linear distortion (e.g., with a 1-bit analog-to-digital converter (ADC)), and different channel conditions (e.g., low-correlated and high-correlated channels). Our work provides novel insights into NN-based channel estimators.
\end{abstract}



\maketitle

\section{INTRODUCTION}
    \IEEEPARstart{C}{urrent} key enabling wireless technologies, such as massive multiple-input-multi-output (MIMO), millimeter-wave (mmWave) communication, and ultra-dense networks have significantly improved the throughput of wireless communications \cite{6736746}. Thanks to the huge bandwidth available in the mmWave band, we can meet the high data rate requirements for future cellular networks, and the high path loss at these frequencies can be compensated by utilizing the beamforming gain of MIMO antennas.
    Nonetheless, these technologies have been mainly studied for half-duplex communication, and the potential of utilizing full-duplex communication has been widely overlooked. In full-duplex transmission, a transceiver can simultaneously transmit and receive over the same carrier frequency, in principle
    doubling the spectral efficiency. Besides, the delay associated with half-duplex transmission is not present in full-duplex systems, 
    thereby enabling the
    low latency requirements of 5G and beyond \cite{6832464}, \cite{9363024}.

    {To realize full-duplex transmission, self-interference (SI) caused by each transmit element over the receive array has to be canceled out.} Consecutive cancellation procedures are usually considered in three domains: 
    propagation, analog, and digital \cite{10163880}. In propagation domain cancellation, SI power is suppressed before reaching the receive chain circuit. Analog domain cancellation is performed before the analog-to-digital converter (ADC) in the receiver analog chain. Finally, digital domain cancellation suppresses the SI power remaining from the propagation and analog domains \cite{6353396}.
    {Implementing full-duplex in mmWave systems poses unique challenges distinct from lower-frequency counterparts. Unlike lower-frequency systems, mmWave transceivers utilize dense antenna arrays and wide bandwidths, necessitating tailored solutions for handling the SI and exploiting spatial degrees of freedom. While lower-frequency full-duplex solutions often rely on analog SI cancellation, mmWave systems require novel approaches due to their unique transceiver architectures and propagation characteristics.}
    

    There are two main antenna configurations used for full-duplex systems: separate and shared. In the separate antenna configuration, separate antennas are dedicated to transmission and reception, yielding high isolation and relatively low interference. In contrast, in the latter configuration, a single antenna is utilized for both transmission and reception, facilitated by the use of circulators to separate the incoming and outgoing signals. This setup offers a simpler overall design and reduced hardware costs, albeit with potential challenges in managing interference and maintaining signal integrity.

    Due to the large number of antenna elements in massive MIMO systems, channel estimation is 
    a complex 
    process with a high pilot overhead. This is the case even in half-duplex
    systems, but is more pronounced in full-duplex systems due
    to the critical need for SI cancellation, which requires estimation of the large SI channel.
     Several studies have been conducted to address channel estimation in full-duplex systems.
    In \cite{6177689}, a channel estimation scheme based on least squares (LS) estimation is proposed for bidirectional communication between a pair of transceivers. The achievable sum rate was analyzed considering the limited dynamic range of the transmitter and receiver and channel estimation error. In \cite{7797501}, an LS-based scheme was utilized for a multi-pair two-way amplify and forward (AF) relaying system. Additionally, a low-pilot overhead scheme was proposed, where two user equipments (UEs) employ the same pilot sequence, while different UE pairs adopt orthogonal pilot sequences.
  
    In \cite{7518612}, an expectation-maximization algorithm is utilized to address the channel estimation problem in a large-scale MIMO system with a full-duplex relay. The authors explore two different scenarios: in the first scenario, both the base station (BS) and the relay estimate their respective individual channels, while in the second scenario, only the BS estimates the cascaded channel between the transmitter, relay, and receiver.
    In \cite{7429783}, channel estimation is performed in both the radio frequency (RF) and baseband domains. In the RF domain, the authors employ the LS technique to estimate the SI channel. 
    Subsequently, a subspace-based algorithm is employed in the baseband to estimate the residual SI and BS-UE channels. The study in \cite{7086821} proposes a specific frame structure for full-duplex systems, for which the achievable rate and SI channel estimation are analyzed. Furthermore, the impact of the 
    calibration period on SI channel estimation is investigated in \cite{7094599}. 

    In \cite{9139277}, the authors have studied the problem of hybrid beamforming with low-resolution phase shifters for full-duplex mmWave large-scale MIMO systems and optimized the sum spectral efficiency (SE) of uplink and downlink UEs, assuming perfect channel state information (CSI) knowledge.
    Two different SI cancellation approaches, namely spatial suppression and SI subtraction, have been compared in \cite{7842132}, taking into account channel estimation errors and channel spatial correlation.
    The uplink and downlink achievable rates have been obtained in \cite{9082190}, considering both the perfect CSI and imperfect CSI cases in full-duplex massive MIMO systems with low-resolution ADCs. 
    A study has been conducted in \cite{8421743} for full-duplex large-scale MIMO cellular systems, considering both non-cooperative and cooperative scenarios. Minimum mean squared error (MMSE) channel estimation is employed for SI and UE channels. Analytical expressions have been derived for the ergodic achievable rate using linear filters such as matched filter (MF) and zero-forcing (ZF).
    Authors in \cite{9456023} have introduced a hybrid beamforming design for mmWave full-duplex, addressing several practical aspects such as codebook-based analog beamforming and beam alignment.

    A neural network (NN) is a specialized tool within the field of machine learning that has demonstrated remarkable effectiveness in solving a wide range of tasks, including computer vision, speech recognition, and autonomous driving \cite{10.5555/3086952}. Thanks to the versatility of NNs, they have also found applications in wireless communication, either in a block-structured manner or in an end-to-end scenario \cite{8839651}. Research in the wireless communication community has explored optimizing traditional processing blocks, such as channel estimation \cite{9410430}, precoding \cite{8618345}, signal detection \cite{9103314}, and CSI feedback \cite{9296555}, using deep learning techniques. Moreover, in \cite{8214233}, an end-to-end communication system has been represented by an autoencoder. In \cite{8644250}, a communication system model is developed without prior information on the channel, with the assistance of conditional generative adversarial networks (GANs).
    

    Numerous studies have focused on NN-based channel estimation. In \cite{8752012}, convolutional neural networks (CNNs) are employed to estimate the mmWave massive MIMO channel, demonstrating that CNNs can achieve superior performance compared to MMSE by effectively exploiting spatial and frequency correlation.
    In \cite{8949757}, the authors utilize a specifically designed CNN called ‘‘deep image prior’’ to denoise the LS estimated massive MIMO channel. Their findings indicate that deep learning can successfully estimate pilot-contaminated massive MIMO channels.
    In another work, \cite{9067011}, fully connected neural networks (FNNs) are utilized to estimate the massive MIMO channel from quantized received measurements with low-resolution ADCs. The study reveals that with sufficiently large antenna arrays, fewer pilots are required for channel estimation.  For beamspace mmWave massive MIMO systems, a learned denoising-based approximate message passing network is used to solve the channel estimation problem \cite{8353153}. Simulation results demonstrate that deep learning-based channel estimation outperforms compressed sensing (CS)-based schemes in this context.

    As highlighted in several papers, such as \cite{6832464}, \cite{9363024}, and \cite{9139277}, the channel estimation problems in full-duplex transmission have not received extensive research attention in the literature. The few existing works mentioned above are primarily focused on simple channel estimation techniques, like LS estimators, which may not ensure high-quality channel estimates in low signal-to-noise ratios (SNRs). On the other hand,  the MMSE channel estimator requires knowledge of the channel correlation matrix and is burdened with high computational complexity.
    {More importantly, classical channel estimation techniques require long pilot dimensions in order to perform well. But as we pointed out, in full-duplex systems, the SI channel is a large MIMO channel, and estimating this channel together with UE channels is extremely costly, rendering the realization of full-duplex systems impractical. Furthermore, interference or residual error from SI and UE signals can heavily disrupt the received pilot signal in full-duplex systems. Another challenge is acquiring the channel from transmit antenna (TX) arrays to the downlink UEs in separate antenna configurations.}

    {In order to circumvent the aforementioned critical problems in channel estimation for full-duplex systems, we examine NNs to estimate both SI and UE channels. The main motivation for adopting NNs for this problem can be summarized as follows: 1) It has been shown in various studies that deep learning can provide remarkably good performance in channel estimation with a short pilot dimension, see e.g., \cite{9067011}, 2) Deep learning has demonstrated impressive performance in noisy conditions \cite{8949757}, holding promise for addressing the problem of heavy interference in the received pilot signal of full-duplex systems. 3) Finally, NNs have been successfully utilized to approximate the mapping of channels in frequency and spatial domains \cite{9048929}. Therefore, NNs are a good candidate for obtaining the channel from TX arrays to downlink UEs by approximating the spatial mapping of the channel from downlink UEs to receive antenna (RX) arrays, to the channel of TX arrays and to downlink UEs.}
    
    
    {
    The role of the number of hidden layers in NN-based channel estimators is not well understood. The general belief in the advantages of depth stems from experiments in machine learning on data modalities such as images or text, which can be far more complex than wireless channels. The latency constraints in wireless systems are also radically different from those in computer vision tasks. As such, the same advantages of depth may not hold for channel estimation. Additionally, the computational complexity and generalization capabilities of deep NNs are significant bottlenecks in implementing NN estimators in practice. Hence, we concentrate on simple NNs ranging from zero to a few hidden layers. We aim to investigate how these simple NNs perform across various channel conditions and system scenarios compared to deep NNs. Additionally, to approximate the channel mapping from TX arrays to downlink UEs to the channel of RX arrays to downlink UEs, we employ FNNs with different numbers of hidden layers.
    }
    
    The summary of the contributions made in this paper is as follows:
    \begin{itemize}
    \item {First, we examine the performance of the NN-based channel estimator with varying numbers of hidden layers and compare them to traditional LS and MMSE techniques, as well as state-of-the-art deep NNs. We consider architectures with 0, 1, 2, and 10 convolutional hidden layers, with the latter corresponding to the deep NN architecture described in \cite{8752012}. To further explore NN-based estimators, we introduce 1-bit ADCs at the full-duplex BS to introduce nonlinear distortion to the received pilot signal. Additionally, to delve deeper into the problem of channel estimation with low-resolution ADCs, we consider cases with 2-bit and 3-bit ADCs. Furthermore, we investigate the performance of NNs under different channel spatial correlation regimes, namely, low-correlation and high-correlation regimes.}
    \item In order to reduce the pilot overhead imposed by full-duplex transmission, we share pilot resources between TX arrays of BS and UEs and explore different pilot dimensions. First, we estimate the SI channel and then cancel out the SI signal from the received pilot signal to estimate the UE channels. 
    We discuss how the SI power influences the channel estimates of UEs when pilot resources are shared between TX arrays and UEs.
    We further analyze the effect of SI cancellation during the pilot phase for UE channel estimation.
    \item  To acquire channel estimates from transmit arrays to downlink UEs in separate antenna configurations, we seek a mapping from the channel of downlink UEs to the RX arrays to the channel of TX arrays to the downlink UEs. Due to the multi-path effect and random scattering environment, there is no simple mathematical formulation for this mapping. Instead, we utilize an NN to approximate it based on the universal approximation \cite{HORNIK1989359}  theory. Accordingly, we estimate the channels of downlink UEs at the RX arrays as well and then map them to the channel from TX arrays to the downlink UEs using NNs. Simulation results show that the NN can learn this mapping successfully.
        {
    
    \item We analyze how our trained models perform when there is a significant distribution shift between the training and test phases. We utilize the DeepMIMO dataset \cite{Alkhateeb2019} to generate channel samples that differ from the data used for training. As expected, the performance of the trained model degrades due to the distribution shift. Moreover, we observe that simpler NNs with fewer hidden layers exhibit better generalization ability compared to deeper NNs, since the number of trainable variables is much lower in shallow NNs. This underscores the advantages of utilizing simpler NN architectures for channel estimation problems.
    }
    \item  Finally, we compare the computational complexity of LS, MMSE, and NN channel estimators in terms of floating point operations (FLOPs). This comparison indicates that 
    the number of FLOPs increases quadratically with the number of received antennas in MMSE estimation, whereas in LS and NN-based estimation, it grows linearly. {
    Furthermore, the comparison of the computational complexity of NNs with different numbers of hidden layers indicates that NNs with a few hidden layers can significantly save computational resources while employing a deep NN poses a major challenge in practice.}

    \end{itemize}
    The rest of this paper is organized as follows. Section \ref{section2} introduces the channel models for SI and UEs, as well as pilot transmission schemes. We examine different channel estimation approaches such as LS and MMSE, and introduce the NN-based estimator for SI and UE channels in Section \ref{section3}, and the RX-TX mapping for separate antenna configurations. In Section \ref{section4}, we provide extensive numerical simulations and compare different channel estimators in terms of normalized mean squared error (NMSE) and computational complexity. Section \ref{section5} concludes the paper.
    
    Throughout the paper the following notations are used:  Matrices are represented using bold uppercase letters, while bold lowercase letters denote vectors. Transpose, transpose conjugate (Hermitian), and matrix inversion operations are denoted by the superscripts $(\cdot)^T$, $(\cdot)^H$, and $(\cdot)^{-1}$, respectively. We use $\mathbb{E}$ to denote expectation, $||\cdot||_{F}$ to represent the Frobenius norm, and $\otimes$ to denote Kronecker product. The identity matrix of size $n \times n$ is represented by $\mathbf{I}_n$. For a generic matrix $\mathbf{A}\in \mathbb{C}^{n\times m}$, $[\mathbf{A}]_{ij}$ refers to the element in the $(i, j)$-th position. The operation $vec(\mathbf{A})$ denotes a column vector obtained by stacking the columns of $\mathbf{A}$ one below the other.
\section{System Model}\label{section2}
\subsection{Channel Modeling}\label{section2_channel_modeling}
\begin{figure}
	\centering
	\includegraphics[height=6.cm,width=8.5cm]{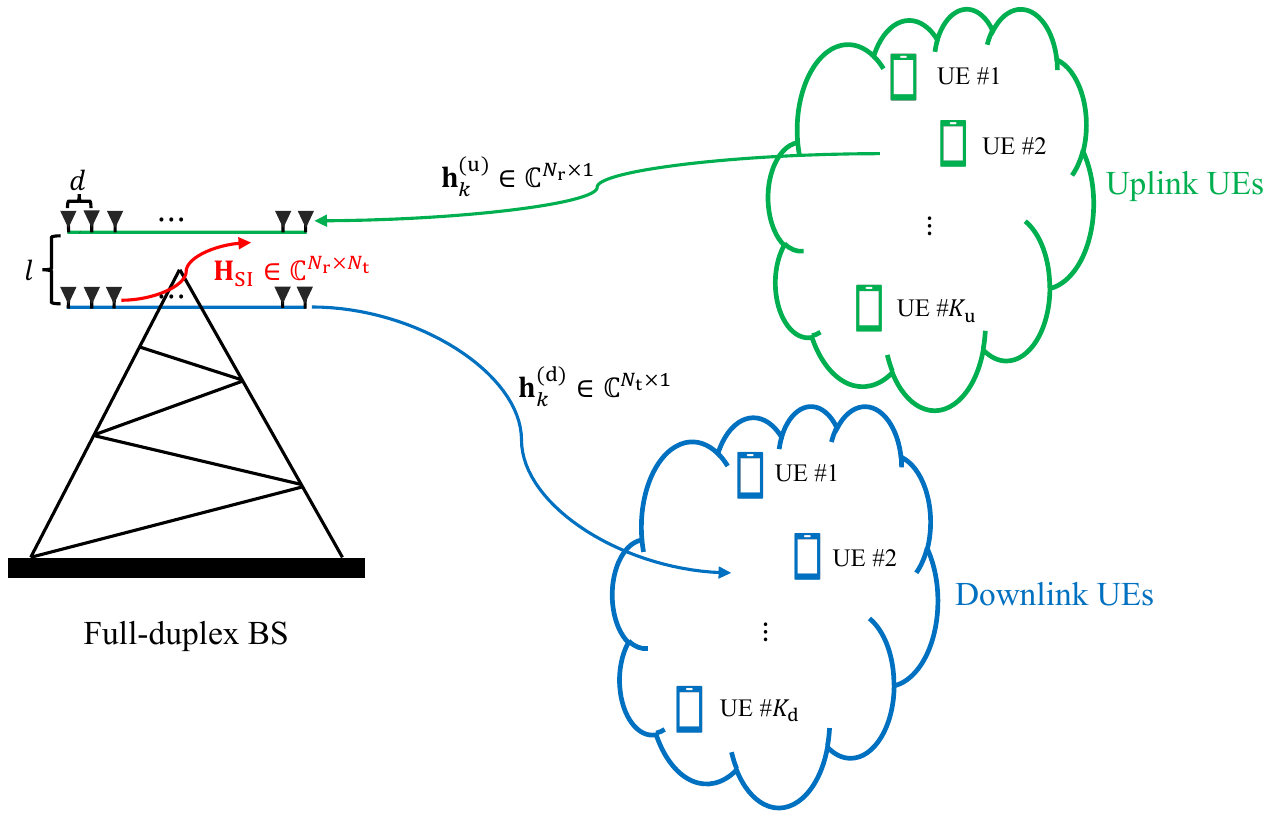}
	\caption{System model with a full-duplex BS serving $ K_{\text{u}} $ single-antenna UEs in the uplink and $K_{\text{d}}$ single-antenna UEs in the downlink channel.}
	\label{fig: Sys_model}
    \end{figure}

    We consider a single-cell system with one full-duplex mmWave  MIMO BS, serving $ K_{\text{u}} $ uplink UEs and $ K_{\text{d}} $ downlink UEs as depicted in the Fig. \ref{fig: Sys_model}.
    Both uplink and downlink UEs are half-duplex devices and the total number of UEs in the cell is $ K=K_{\text{u}}+K_{\text{d}} $. We consider separate antenna configurations, where the BS is equipped with two uniform linear arrays to enable full-duplex transmission. The number of transmit and receive antennas are $ N_{\text{t}} $ and $ N_{\text{r}} $, respectively, and they are linearly placed with a uniform distance.
    {
    
    In order to fully investigate different estimators within various scenarios, in this work, the UE channels follow a geometrical channel model as follows \cite{6847111}:
    \begin{equation}
	\mathbf h_{k}^{(\text{u})} = \sqrt{\frac{N_{\text{r}}}{P}} \sqrt{\beta_{\text{u},k}} \sum_{i=0}^{P-1}\alpha_{i,k} \mathbf{a}_{\text{a}}(\theta_{i,k}), \;\;\; k=1,2,\dots,K_{\text{u}},
    \end{equation}
    \begin{equation}
	\mathbf h_{k}^{(\text{d})} = \sqrt{\frac{N_{\text{t}}}{P}} \sqrt{\beta_{\text{d},k}} \sum_{i=0}^{P-1} \alpha_{i,k} \mathbf{a}_{\text{d}}(\theta_{i,k}), \;\;\; k=1,2,\dots,K_{\text{d}},
    \end{equation}
    where $\mathbf h_{k}^{(\text{u})}$($\mathbf h_{k}^{(\text{d})}$) is the uplink (downlink) channel between the $k$-th uplink (downlink) UE and receive (transmit) antenna arrays at BS. Furthermore, 
    $P$ represents the number of multi-path components,  $\alpha_{i,k} \sim \mathcal{CN}(0,1)$ are the small-scale fading factors of individual paths, and $ \beta_{\text{u}(\text{d}),k} $ stands for the large-scale fading 
    \begin{equation}
	\beta_{\text{u}(\text{d}),k}= \Gamma- 10\eta \text{log}(r_{\text{u}(\text{d}),k})+\chi_k,
    \end{equation}
    where $ \Gamma $ denotes the average channel gain in dB at a reference distance of 1 m and specific carrier frequency, $ r_{\text{u}(\text{d}),k} $ is the uplink (downlink) distance between transmitter and receiver, $ \chi_k $ is the shadow fading with a lognormal distribution $\mathcal{LN}(0,\sigma_{\text{sf}})$, where $ \sigma_{\text{sf}}$ is standard deviation, and $ \eta $ is the path loss exponent. Furthermore, $\mathbf{a}_{\text{a}}(\theta)$ and $\mathbf{a}_{\text{d}}(\theta)$ are, respectively, the uniform linear array (ULA) responses of the receive and transmit antennas, given by
    \begin{equation}
    \begin{split}
    \mathbf{a}_{\text{a}}(\theta) &= \sqrt{\dfrac{1}{N_{\text{r}}}} \left[1, \exp\left(j2\pi \frac{d}{\lambda}(\sin\theta)\right), \ldots,\right.\\
    &\left. \exp\left(j2\pi \frac{d}{\lambda}(N_{\text{r}}-1)\sin\theta\right) \right]^T,
    \end{split}
    \end{equation}
and
    \begin{equation}
    \begin{split}
    \mathbf{a}_{\text{a}}(\theta) &= \sqrt{\dfrac{1}{N_{\text{t}}}} \left[1, \exp\left(j2\pi \frac{d}{\lambda}(\sin\theta)\right), \ldots,\right.\\
    &\left. \exp\left(j2\pi \frac{d}{\lambda}(N_{\text{t}}-1)\sin\theta\right) \right]^T,
    \end{split}
    \end{equation}
    }
    where $ \theta $ is the angle of arrival/departure (AoA/AoD) to/from the receive/transmit antenna arrays at the BS, and $d$  and $ \lambda $ are the antenna spacing and wavelength, respectively.  We assume that the AoA and AoD follow the local scattering model with a uniform distribution $\left[-\frac{\theta_{\text{AS}}}{2}, \frac{\theta_{\text{AS}}}{2}\right]$ \cite{massivemimobook}, where $\theta_{\text{AS}}$ is the angular spread (AS) of the multi-path components. 
    {
    
    In the full-duplex transmission, there is cross-link interference between uplink and downlink UEs that needs to be carefully modeled. Nonetheless, since the focus of the paper is channel estimation using uplink pilots, we are neglecting the cross-link interference in our study.
    }
    
    We denote the SI channel from the transmit antenna arrays to the receive antenna arrays as $ \mathbf{H}_{\text{SI}} $, which is modeled as a Rician fading channel due to the presence of a strong line-of-sight (LoS) path.  Hence, the SI channel consists of two components: the near-field and the far-field terms \cite{6832464}: 
    \begin{equation}\label{eq:channel_phys}
	\mathbf{H}_{\text{SI}} =  \sqrt{\epsilon_{\text{SI}}}\sqrt{\frac{\kappa}{\kappa+1}}\mathbf{H}_{\text{SI,NF}} + \sqrt{\frac{1}{\kappa+1}}\mathbf{H}_{\text{SI,FF}}.
    \end{equation}

    {
    
    We assume that SI is sufficiently suppressed in the propagation domain and the received signal at the baseband does not exceed the dynamic range of the ADCs. We refer to the propagation domain SI suppression factor as $ \epsilon_{\text{SI}} $.
    Note that in the propagation domain, only the near field channel from the transmit antenna arrays to the receive antenna arrays can be suppressed. In the digital domain, the residual near-field and far-field SI channels can be further suppressed after estimating the SI channel. 
    }
    The near-field term $\mathbf{H}_{\text{SI,NF}}$ arises from the close-in placement of the transmit and receive antennas, while the far-field term $\mathbf{H}_{\text{SI,FF}}$ is a result of the reflections in the environment. Here, $\kappa$ is the Rician factor and $\mathbf{H}_{\text{SI,FF}}$ follows a similar geometrical channel modeling as the UE channels \cite{6847111}:
    {
    
    \begin{equation}
	\mathbf{H}_{\text{SI,FF}} = \sqrt{\frac{N_{\text{t}}N_{\text{r}}}{P}} \sqrt{\beta_{\text{SI}}} \sum_{i=0}^{P-1} \alpha_{i} \mathbf{a_{\text{a}}} (\theta_{i})\mathbf{a_{\text{d}}}^{T}(\theta_{i}),
    \end{equation}
    }
    and $\beta_{\text{SI}}$ follows a similar formulation to $\beta_{\text{u}(\text{d}),k}$.
    
    The near-field term does not satisfy the far-field condition and therefore requires a completely different channel modeling approach \cite{10176267}, \cite{9794906}. In this paper, we adopt the 
    model
    in \cite{8246856}, given by
    \begin{equation}
	\left[\mathbf{H_{\text{SI,NF}}} \right]_{nm} = \frac{\rho}{r_{nm}}\exp\left(-2\pi j \frac{r_{nm}}{\lambda}\right),
    \end{equation}
    where $r_{nm}$ is the distance between the $n$-th receive antenna and the $m$-th transmit antenna, and $\rho$ is a constant {for power normalization.}
\subsection{Pilot transmission schemes}\label{Pilot_transmission_schemes}
    We consider various pilot dimensions for channel estimation in full-duplex MIMO systems. Since uplink and downlink transmissions happen at the same time and frequency in full-duplex systems, only uplink pilots will be transmitted from UEs. To estimate the channels of UEs and SI, both downlink and uplink UEs, along with transmit antennas at the BS, transmit pilot signals to the receive antennas at the BS. Specifically, we assume that all the $K$ UEs transmit their pilot signals to the RX arrays in the channel estimation phase, whereas in the payload data transmission phase, a subset of $K_{\text{u}}$ UEs will be in the uplink mode and $K_{\text{d}}$ UEs will be in the downlink mode.
    
\begin{figure}
	\centering
	\includegraphics[height=6.cm,width=8.5cm]{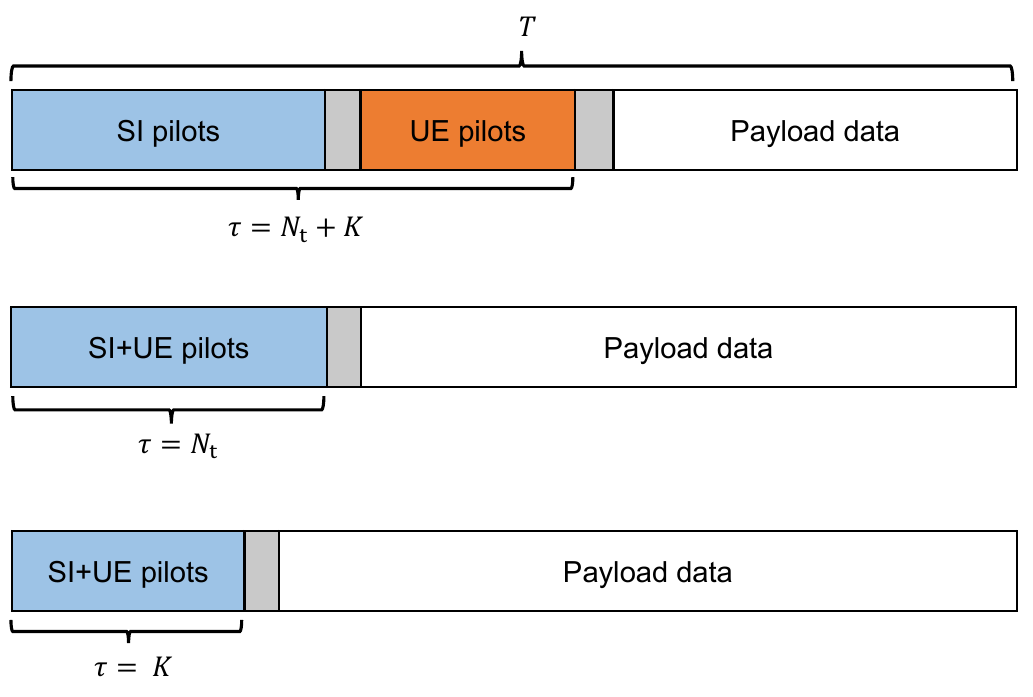}
	\caption{Different pilot dimensions. $T$ is the frame dimension during one coherence block and $\tau$ is the pilot dimension for estimating SI and UE channels.}
	\label{fig:Pilots_fig}
    \end{figure}

    In the baseline scenario, we assume UEs and transmit antennas of the BS send their pilot signals at different pilot resources. Due to large antenna arrays, with this pilot transmission, the pilot overhead will be extremely high.
    In the second scenario, we assume the available pilot resources will be equal to the number of transmit antennas at the BS, and UEs will reuse the same pilot resources utilized by the BS to estimate BS-UE channels. Compared to the first scenario, the pilot overhead will be lower, but the received pilot signals will be contaminated by UEs/TX arrays when estimating SI/UE channels.
    To further reduce the pilot overhead, we reduce the number of pilots to the number of UEs. In this case, besides interference, the SI channel must be estimated with fewer pilots than the channel dimension, but with some performance loss, we can achieve a comparable pilot overhead as half-duplex systems. Three different pilot dimensions considered for estimating SI and UE channels are depicted in Fig. \ref{fig:Pilots_fig}.

We assume that the received SI signal power remains stronger than the received UE signal power, even after SI cancellation in the propagation domain. Hence, our approach involves initially estimating the SI channel and subsequently subtracting the estimated SI signal from the received pilot signal to estimate the UE channels.
\subsection{SI channel estimation}

 For SI channel estimation, the received pilot signal at the RX arrays can be written as
\begin{equation}\label{Y_SI}
\mathbf{Y_{\text{SI}}} = \sqrt{{\mathrm{SNR}_{\text{SI}}}}\mathbf{H_{\text{SI}}}\mathbf{F}\mathbf{X_{\text{SI}}} + \sqrt{{\mathrm{SNR}_{\text{UE}}}}\mathbf{H_{\text{UE}}}\mathbf{X_{\text{UE}}} + \mathbf{N},
\end{equation}
where $\mathbf{Y_{\text{SI}}} \in \mathbb{C}^{N_{\text{r}}\times \tau} $ is the received pilot signal at receive antennas of BS during $\tau$ pilot transmissions.  $\mathbf{X_{\text{UE}}}\in \mathbb{C}^{K\times \tau}$ is the transmitted pilot signal from the $K$ UEs,  $\mathbf{X_{\text{SI}}} \in \mathbb{C}^{\tau\times \tau} $ is a diagonal matrix  whose diagonal elements are transmitted pilot signals from the transmit antennas at BS,  $\mathbf{F}\in \mathbb{C}^{N_{\text{t}}\times \tau}$ is the precoding matrix of transmit antennas of BS, and $  \mathbf{H_\text{UE}} = \left[\mathbf{h}_{1}^{(\text{u})},\mathbf{h}_{2}^{(\text{u})},\dots,\mathbf{h}_{K_{\text{u}}}^{(\text{u})},\mathbf{h}_{1}^{(\text{d})},\mathbf{h}_{2}^{(\text{d})},\dots,\mathbf{h}_{K_{\text{d}}}^{(\text{d})}\right]\in \mathbb{C}^{N_{\text{r}}\times K} $ is the concatenated uplink and downlink channel response matrix, where the $k$-th column corresponds to the channel between the $k$-th uplink or downlink UE and the receive antenna at the BS. {In this paper, we assume fully digital beamforming at the full-duplex BS, and considering analog or hybrid precoding is left for future work.} 
We assume all UEs have the same transmit power $p_{\text{UE}}$, and $p_{\text{SI}}$ denotes the transmit power from transmit antenna arrays. We define
$\mathrm{SNR}_{\text{SI}}$ and $\mathrm{SNR}_{\text{UE}}$ as
{

\begin{equation}
    \mathrm{SNR}_{\text{SI}} = \frac{p_{\text{SI}}\abs{\mathbf{H_{\text{SI}}}}^2}{\sigma_\text{n}^2},
\end{equation}
\begin{equation}
    \mathrm{SNR}_{\text{UE}} = \frac{p_{\text{UE}}\abs{\mathbf{H_{\text{UE}}}}^2}{\sigma_\text{n}^2},
\end{equation}
}
where $\sigma_\text{n}^2$ is the noise variance.
The matrix $\mathbf{N}\in \mathbb{C}^{N_\text{r}\times \tau}$ is the additive noise comprised of independent and identically distributed (i.i.d.) elements following a zero-mean unit variance Gaussian distribution.

Note that when the pilot dimension is $\tau = N_{\text{t}}+K$, the SI and UE channel estimation happens at orthogonal pilot resources. Hence, the received pilot signal for SI channel estimation does not interfere with the pilots transmitted by UEs.

During pilot transmission,  we apply an orthogonal pilot codebook for the precoded SI signals from the transmit antenna arrays and pilot signals from UEs. Specifically, we utilize the discrete Fourier transform (DFT) pilot codebook, given by $\mathbf{F}\mathbf{X_{\text{SI}}} = \mathbf{W}_{N_{\text{t}}\times\tau}, \mathbf{X_{\text{UE}}} = \mathbf{W}_{K\times\tau}$, where,
\begin{equation}
\begin{split}
    &\mathbf{W}_{M\times \tau} =\\
    &\frac{1}{\sqrt{M}}     
    \begin{bmatrix}
    1&1&1&\cdots &1 \\
    1&\omega&\omega^2&\cdots&\omega^{\tau-1} \\
    1&\omega^2&\omega^4&\cdots&\omega^{2(\tau-1)}\\ 
    1&\omega^3&\omega^6&\cdots&\omega^{3(\tau-1)}\\
    \vdots&\vdots&\vdots&\ddots&\vdots\\
    1&\omega^{M-1}&\omega^{2(M-1)}&\cdots&\omega^{(M-1)(\tau-1)}
    \end{bmatrix}
\end{split}
\end{equation}
and $\omega = e^{-2\pi j/\tau}$.

\subsection{UE channel estimation}
To estimate the UE channels, we use the estimated SI channel and cancel out the SI signal power from the received pilot signal.  Therefore, the received pilot signal for UE channel estimation becomes
\begin{equation}\label{UE_received_pilot}
\begin{split}
    \mathbf{Y_{\text{UE}}} =& \sqrt{{\mathrm{SNR}_{\text{UE}}}}\mathbf{H_{\text{UE}}}\mathbf{X_{\text{UE}}} + \sqrt{{\mathrm{SNR}_{\text{SI}}}}\mathbf{E_{\text{SI}}}\mathbf{F}\mathbf{X_{\text{SI}}} + \mathbf{N},
\end{split}
\end{equation}
where $\mathbf{E_{\text{SI}}} \triangleq \mathbf{H_{\text{SI}}}-\mathbf{\hat{H}_{\text{SI}}}$  is the SI cancelation error and $\mathbf{\hat{H}_{\text{SI}}}$ is the estimated SI channel. 

Similarly, for the pilot dimension $\tau=N_{\text{t}}+K$, the received pilot signal does not contain the signal from the SI channel.

Typically, the channels of nearby UEs are correlated, and for the pilot dimension $\tau=N_{\text{t}}$ and $\tau=K$, the residual error of SI cancellation introduces further correlation in the received pilot signals used for UE channel estimation. We estimate the entire UE channel matrix $\mathbf{H_{\text{UE}}}$. In this way, we take advantage of the correlation resulting from imperfect SI cancellation and leverage the additional information for more accurate estimation of the UE channels.


\section{LS, MMSE, NN-based Channel Estimation}\label{section3}
In this section, we present three channel estimation methods: LS, MMSE, and NN-based estimators for both SI and UE channels. First, we will recall the LS  and MMSE  channel estimators. Subsequently, we present channel estimation using NNs for estimating the SI and UE channels. For SI channel estimation, we correlate the received pilot signal with the pilot matrix transmitted from transmit antenna arrays, therefore, we define
\begin{equation}
    \mathbf{\hat{Y}_{\text{SI}}} = \mathbf{Y_{\text{SI}}} \mathbf{W}^H_{N_{\text{t}}\times \tau}.
\end{equation}
Similarly for the UE channel estimation, we have
\begin{equation}
    \mathbf{\hat{Y}_{\text{UE}}} = \mathbf{Y_{\text{UE}}} \mathbf{W}^H_{K\times \tau}.
\end{equation}
\subsection{LS channel estimator}
The LS channel estimator is the most simple channel estimation technique that does not need any prior knowledge about channel statistics and finds the channel coefficients that minimize the mean square error (MSE) between the estimated channel and the received pilot signal. The LS channel estimator can be  calculated as follows
\begin{equation}
    \mathbf{\hat{H}}_{q,\text{LS}} = \frac{1}{\tau \sqrt{{\mathrm{SNR}_{\text{q}}}}}\mathbf{\hat{Y}}_{q},
\end{equation}
where $q\in \{\text{SI}, \text{UE}\}$.


\subsection{MMSE channel estimator}
MMSE is a Bayesian estimator that aims to minimize the MSE between the estimated channel and the true channel. It offers improved performance compared to the LS estimator by taking into account the statistical properties of the channel and noise. To derive the MMSE channel estimator, we rewrite the received pilot signal for SI and UE channel estimation in a vector form
\begin{equation}
\begin{split}
    vec(\mathbf{Y_{\text{SI}}}) =& \sqrt{\mathrm{SNR}_{\text{SI}}}\mathbf{\Tilde{X}_{\text{SI}}}vec(\mathbf{H_{\text{SI}}}) + \\&
    \sqrt{\mathrm{SNR}_{\text{UE}}}\mathbf{\Tilde{X}_{\text{UE}}}vec(\mathbf{H_{\text{UE}}}) + vec(\mathbf{N}),
\end{split}
\end{equation}

\begin{equation}
\begin{split}
    vec(\mathbf{Y_{\text{UE}}}) =& \sqrt{\mathrm{SNR}_{\text{UE}}}\mathbf{\Tilde{X}_{\text{UE}}}vec(\mathbf{H_{\text{UE}}}) + \\&
    \sqrt{\mathrm{SNR}_{\text{SI}}}\mathbf{\Tilde{X}_{\text{SI}}}vec(\mathbf{E_{\text{SI}}}) + vec(\mathbf{N}),
\end{split}
\end{equation}
where $\mathbf{\Tilde{X}_{\text{SI}}} = \mathbf{(FX_{\text{SI}})}^{T}\otimes \mathbf{I_{N_{\text{r}}}}$ and $\mathbf{\Tilde{X}_{\text{UE}}} = \mathbf{(X_{\text{UE}})}^{T}\otimes \mathbf{I_{N_{\text{r}}}}$.

The channel covariance matrix is defined as follows
\begin{equation}
    \mathbf{R}_{q} \triangleq \mathbb{E}\left[vec(\mathbf{H}_q)vec(\mathbf{H}_q)^H\right].
\end{equation}

The channel covariance matrix for the MIMO channel can be related to the transmit and receive covariance matrix via the Kronecker product
\begin{equation}
    \mathbf{R} = \mathbf{R_{\text{t}}}\otimes \mathbf{\mathbf{R_{\text{r}}}},
\end{equation}
where $\mathbf{R_{\text{t}}}$ and $\mathbf{R_{\text{r}}}$ are the transmit and receive covariance matrices.

The MMSE channel estimator for $\mathbf{H}_q$, $\mathbf{\hat{H}}_{q,\text{MMSE}}$, minimizes the mean square error (MSE) $\mathbb{E}\left[||\mathbf{H}_q - \mathbf{\hat{H}}_{q,\text{MMSE}}||^2\right]$. We have
\begin{equation}
    vec(\mathbf{\hat{H}}_{q,\text{MMSE}}) = \mathbf{R}_{\mathbf{H}_q\mathbf{Y}_q}\mathbf{R}_{\mathbf{Y}_q}^{-1}vec(\mathbf{Y}_q),
\end{equation}
where
\begin{equation}
   \mathbf{R}_{\mathbf{H}_q\mathbf{Y}_q} \triangleq \mathbb{E}\left[vec(\mathbf{H}_q)vec(\mathbf{Y}_q)^H\right],
\end{equation}
\begin{equation}
    \mathbf{R}_{\mathbf{Y}_q} \triangleq \mathbb{E}\left[vec(\mathbf{Y}_q)vec(\mathbf{Y}_q)^H\right].
\end{equation}
We can formulate the MMSE estimates of SI and UE channels as in equations \eqref{MMSE_SI} and \eqref{MMSE_UE} provided at the top of the next page, respectively. In \eqref{MMSE_UE}, $\mathbf{R_{\text{E}}}$ represents the error covariance matrix of SI cancellation, which can be calculated as in \eqref{Error_cov}.
\setlength{\floatsep}{12pt plus 2pt minus 2pt}
\setlength{\textfloatsep}{20pt plus 2pt minus 4pt}
\setlength{\intextsep}{14pt plus 4pt minus 4pt}

\begin{figure*}[h!]
    \begin{equation}\label{MMSE_SI}
    vec(\mathbf{\hat{H}}_{\text{SI,MMSE}}) = \sqrt{\mathrm{SNR}_{\text{SI}}}\mathbf{R_{\text{SI}}}\mathbf{\Tilde{X}_{\text{SI}}}^H
   \left(\mathrm{SNR}_{\text{SI}}\mathbf{\Tilde{X}_{\text{SI}}} \mathbf{R_{\text{SI}}} \mathbf{\Tilde{X}_{\text{SI}}}^H + \mathrm{SNR}_{\text{UE}}\mathbf{\Tilde{X}_{\text{UE}}} \mathbf{R_{\text{UE}}} \mathbf{\Tilde{X}_{\text{UE}}}^H + \mathbf{I}_{N_\text{t}N_\text{r}}  \right)^{-1} vec(\mathbf{Y_{\text{SI}}}).
    \end{equation}
\end{figure*}
\begin{figure*}[h!]
    \begin{equation}\label{MMSE_UE}
    vec(\mathbf{\hat{H}}_{\text{UE,MMSE}}) = \sqrt{\mathrm{SNR}_{\text{UE}}}\mathbf{R_{\text{UE}}}\mathbf{\Tilde{X}_{\text{UE}}}^H
   \left(\mathrm{SNR}_{\text{UE}}\mathbf{\Tilde{X}_{\text{UE}}} \mathbf{R_{\text{UE}}} \mathbf{\Tilde{X}_{\text{UE}}}^H + \mathrm{SNR}_{\text{SI}}\mathbf{\Tilde{X}_{\text{SI}}} \mathbf{R_{\text{E}}} \mathbf{\Tilde{X}_{\text{UE}}}^H +\mathbf{I}_{N_{\text{t}}N_\text{r}}  \right)^{-1} vec(\mathbf{Y_{\text{UE}}}).
    \end{equation}
\end{figure*}
\begin{figure*}[h!]
    \begin{equation}\label{Error_cov}
    \mathbf{R_\text{E}} = \mathbf{R_\text{SI}}  - \mathrm{SNR}_{\text{SI}}\mathbf{R_{\text{SI}}} \mathbf{\Tilde{X}_{\text{SI}}}^H \left(\mathrm{SNR}_{\text{SI}}\mathbf{\Tilde{X}_{\text{SI}}} \mathbf{R_{\text{SI}}} \mathbf{\Tilde{X}_{\text{SI}}}^H + \mathrm{SNR}_{\text{UE}}\mathbf{\Tilde{X}_{\text{UE}}} \mathbf{R_{\text{UE}}} \mathbf{\Tilde{X}_{\text{UE}}}^H + \mathbf{I}_{N_\text{t}N_\text{r}}  \right)^{-1}\mathbf{\Tilde{X}_{\text{SI}}} \mathbf{R_{\text{SI}}}.
    \end{equation}
\end{figure*}

Note that for pilot dimension $\tau = N_{\text{t}} +K$, the MMSE estimates of the SI and UE channels simplify to the following equations using Woodbury matrix identity \cite{5199136},
\begin{equation}
\begin{split}
    &vec(\mathbf{\hat{H}_{\text{SI,MMSE}}}) =\\& \sqrt{\mathrm{SNR}_{\text{SI}}}
   \left(\mathbf{R_{\text{SI}}}^{-1} +\tau \mathrm{SNR}_{\text{SI}}  \mathbf{I}_{N_{\text{t}}N_{\text{r}}}  \right)^{-1} \mathbf{\Tilde{X}_{\text{SI}}}^Hvec(\mathbf{Y_{\text{SI}}}),
\end{split}
\end{equation}
\begin{equation}
\begin{split}
    &vec(\mathbf{\hat{H}_{\text{UE,MMSE}}}) =\\& \sqrt{\mathrm{SNR}_{\text{UE}}}
   \left(\mathbf{R_{\text{SI}}}^{-1} +\tau \mathrm{SNR}_{\text{UE}} \mathbf{I}_{N_{\text{t}}N_{\text{r}}}  \right)^{-1} \mathbf{\Tilde{X}_{\text{SI}}}^Hvec(\mathbf{Y_{\text{SI}}}).
\end{split}
\end{equation}



\subsection{NN channel estimator}\label{DNN channel estimator}
Although conventional channel estimation techniques such as MMSE can leverage the spatial correlation present in large antenna arrays, they require prior knowledge of the channel covariance matrix and are computationally intensive due to matrix inversion involving a large matrix. In various fields of research, it has been demonstrated that NNs can utilize the inherent structure of data by solely observing a sufficient number of data points. Different network architectures, such as FNNs, CNNs, recurrent neural networks (RNNs), etc., can be utilized depending on the characteristics of the data. For instance, CNNs are suitable for addressing problems with feature correlation and local dependencies among them, while RNNs are more effective for handling memory-based features like time series \cite{10.5555/3086952}.

It has been shown in various studies that the wireless channel with large antenna arrays is spatially correlated, see e.g., \cite{8861014}. To exploit this spatial correlation for channel estimation, we employ CNNs, which are suitable for capturing the local dependency of features among data points. 
{

The overall architecture of CNN for SI and UE channel estimation is depicted in Fig. \ref{CNN_architecture}.
\begin{figure*}
	\centering
	\includegraphics[scale=0.5]{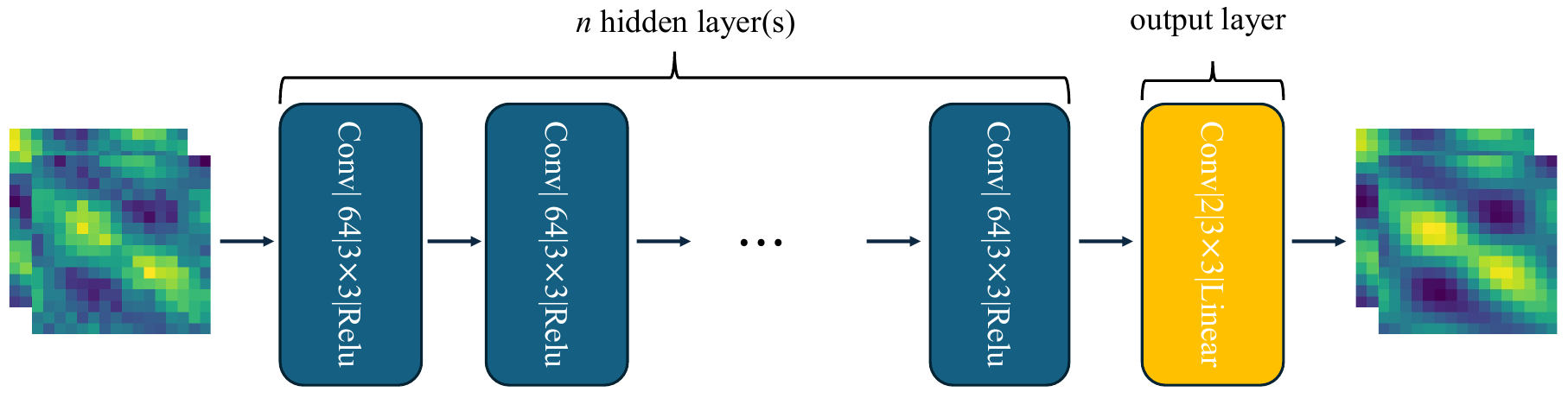}
	\caption{{The CNN architecture employed for SI and UE channel estimation. we apply $n$  ($n = 0, 1, 2, 10$) convolutional hidden layer(s) each with a $3\times 3$ window size and 64 convolutional kernels.}}
        \label{CNN_architecture}
\end{figure*}
}

The input to the CNN is the correlated pilot signal, i.e., $\mathbf{\hat{Y}}_{q}$, and the output is the estimated channel, $\mathbf{\hat{H}}_{q, \text{NN}}$, $q\in \{\text{SI}, \text{UE}\}$. In convolutional layers, we apply a $3\times 3$ window size sliding through the whole input features with a unit stride size. Different numbers of hidden layers are employed, where each layer applies 64 convolutional kernels to extract features from the successive windows of its input features. The effect of the number of hidden layers will be carefully examined in Section \ref{section4}.

To keep the dimensions of the output and input fixed, we utilize padding after convolution processing. We apply rectified linear unit (ReLU) as the activation function for the hidden layers, while for the output layer, linear activation is used. Since tensors do not support complex operations, the input to the CNN is converted to three-dimensional tensors, where the third dimension stores the real and imaginary parts of the complex data samples. Therefore, if we define $\mathbf{X_{\text{tr}}}$ and $\mathbf{Y_{\text{tr}}}$ as the input and labels of the CNN during training, we have
\begin{equation}
\begin{split}
    &\mathbf{X_{\text{tr}}}\left[:,:,0\right] \triangleq \Re\left\{\mathbf{\hat{Y}}_{q}\right\},\\
    &\mathbf{X_{\text{tr}}}\left[:,:,1\right] \triangleq \Im\left\{\mathbf{\hat{Y}}_{q}\right\}.
\end{split}
\end{equation}
\begin{equation}
\begin{split}
    &\mathbf{Y_{\text{tr}}}\left[:,:,0\right] \triangleq \Re\left\{\mathbf{H}_{q}\right\},\\
    &\mathbf{Y_{\text{tr}}}\left[:,:,1\right] \triangleq \Im\left\{\mathbf{H}_{q}\right\}.
\end{split}
\end{equation}

{For training, a dataset consisting of $M_{\text{tr}}$ samples is generated, with $(\mathbf{\hat{Y}}_{q}^{(n)}, \mathbf{H}_{q}^{(n)})$ representing the $n$-th sample. We employ supervised learning, where $\mathbf{H}_{q}$ is regarded as the label during training. To obtain the ground truth channel samples, $\mathbf{H}_{q}^{(n)}$, for SI and UE channels, we follow the channel model introduced in Section \ref{section2}. Each realization of data samples in the dataset is generated based on this channel model, where, in each realization, the small-scale fading coefficient, large-scale fading, and AoA/AoD take on a new realization based on their corresponding distributions. Specifically, we draw samples from $\mathcal{CN}(0,1)$ for every multi-path component for small-scale fading, $\mathcal{U}\left[-\frac{\theta_{\text{AS}}}{2}, \frac{\theta_{\text{AS}}}{2}\right]$ for every multi-path component for AoA/AoD, and $\mathcal{LN}(0,\sigma_{sf})$ for shadow fading in large-scale fading. After generating channel realizations, the received pilot signal for SI and UE channel estimation, $\mathbf{\hat{Y}}_{q}^{(n)}$, is generated by adding Gaussian noise to every realization and multiplying by the pilot matrix corresponding to different pilot dimensions.}
 We apply min-max scaling to scale the dataset in the range $(0,1)$. Such normalization is highly recommended for training machine learning models. We consider MSE as the loss function
\begin{equation}
    \text{MSE} = \frac{1}{M_{\text{tr}}}\sum_{i=1}^{M_{\text{tr}}} ||\mathbf{X_{\text{tr}}} - \mathbf{Y_{\text{tr}}}||_{F}^{2}.
\end{equation}

 \begin{figure}[!t]
	\centering
	\includegraphics[height=6.5cm,width=7cm]{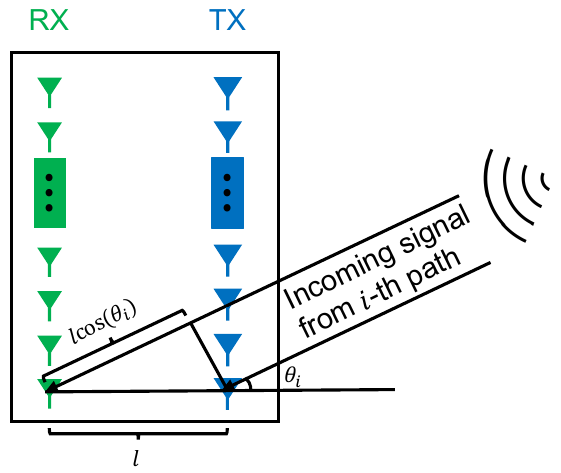}
	\caption{Phase shift between the RX and TX arrays in the separate antenna configuration in a full-duplex BS.}\vspace{-5mm}
	\label{RX_TX}
\end{figure}

\subsection{RX-TX channel mapping}
Since in separate antenna configurations, two different antenna arrays are utilized for uplink reception and downlink transmission, uplink and downlink channels experience different channel realizations. So far we have assumed that RX arrays at BS receive pilot signals from both uplink and downlink UEs and estimate all channels from both uplink and downlink UEs to the RX arrays. For downlink transmission, the full-duplex BS needs the CSI from the TX arrays to downlink UEs. The TX arrays are not capable of receiving and processing pilot signals and implementing a separate receive RF chain and ADC at the TX arrays for just pilot processing will be costly. As a result, we need to somehow map the channel from the downlink UEs to the RX arrays to the channel from the TX arrays to the downlink UEs. Due to the multi-path effect and random scattering environment, it is not straightforward to derive mathematically the relation of such mapping. More specifically, we assume that the RX-UE channel is
\begin{equation}
    \mathbf{h_{\text{UE,RX}}} = \sqrt{\frac{1}{N_{\text{r}}}} \sqrt{\beta_{\text{u}}} \sum_{i=0}^{P-1} \alpha_{i}\mathbf{a}_{\text{a}}(\theta_{i}),
\end{equation}
where $ a_{i}$ and $\theta_{i}$  are the amplitude and the AoA of $ i $-th path, respectively. As shown in Fig. \ref{RX_TX}, TX and RX arrays are at a close distance in the order of a few wavelengths from each other; Therefore, the channel amplitude for each path is essentially the same for both the RX and TX arrays, while the antenna separation creates a delay depending on the AoA of each path. Thus, the TX-UE channel will be
\begin{equation}
    \mathbf{h_{\text{UE,TX}}} = \sqrt{\frac{1}{N_{\text{t}}}} \sqrt{\beta_{\text{d}}} \sum_{i=0}^{P-1} \alpha_{i}\mathbf{a}_{\text{d}}({\theta_{i})} e^{-j\frac{2\pi}{\lambda}lcos\theta_i}.
\end{equation}

The mathematical relationship between $\mathbf{h_{\text{UE,TX}}}$ and $\mathbf{h_{\text{UE,RX}}}$ is not well-defined due to the random AoA and the effects of multi-path propagation. On the other hand, based on universal approximation theory \cite{HORNIK1989359}, feedforward NNs are capable of approximating any continuous function. This theory suggests that, given enough computational resources and data, it is possible to build NN models that can accurately approximate any function. Therefore, universal 
approximation theory inspires us to employ NNs to map the channel from the downlink UEs to the RX array to the channel from the TX array to the downlink UEs. Furthermore, recent studies \cite{9048929} have demonstrated the existence of a spatial mapping function that can effectively map the channel from one set of antenna arrays to another.
To accomplish this, we utilize an FNN with varying numbers of hidden layers. The input for FNN is the channel from the downlink UEs to the RX array, while the output is the channel from the TX array to the downlink UEs. Therefore, for the training of FNN, we collect $M_{\text{tr}}$ samples of $\mathbf{h_{\text{UE,RX}}}$ as the input and $\mathbf{h_{\text{UE,TX}}}$ as the corresponding label.
{To generate data samples for RX-TX channel mapping, we first generate the ground truth channel realizations for $\mathbf{h_{\text{UE,RX}}}$, and their corresponding labels for $\mathbf{h_{\text{UE,TX}}}$ are generated by adding its corresponding delay to each multi-path component.
}
 Again, we apply min-max normalization and MSE loss function for training. We create the following data samples as the input and label of FNN to work with real-valued tensors:
\begin{equation}
\begin{split}
    &\mathbf{x_{\text{tr}}}[0:N_{\text{r}}] = \Re\left\{\mathbf{h_{\text{UE,RX}}}\right\},\\
    &\mathbf{x_{\text{tr}}}[N_{\text{r}}:2 N_{\text{r}}] = \Im\left\{\mathbf{h_{\text{UE,RX}}}\right\}.
\end{split}
\end{equation}
\begin{equation}
\begin{split}
    &\mathbf{y_{\text{tr}}}[0:N_{\text{t}}] = \Re\left\{\mathbf{h_{\text{UE,TX}}}\right\},\\
    &\mathbf{y_{\text{tr}}}[N_{\text{t}}:2 N_{\text{t}}] = \Im\left\{\mathbf{h_{\text{UE,TX}}}\right\}.
\end{split}
\end{equation}

  \section{Simulation results}\label{section4}
In this section, we present our simulation results for SI and UE channel estimation. We compare channel estimators that we introduced in previous sections, i.e., LS, MMSE, and NN for different pilot dimensions. The NMSE is considered the performance metric to compare the different channel estimators, and it is defined as 
\begin{equation}
    \text{NMSE} \triangleq \mathbb{E}\left[ \frac{||\mathbf{H_\text{true}} - \mathbf{H_\text{est}}||_{F}^{2}}{||\mathbf{H_\text{true}}||_{F}^{2}}\right],
\end{equation}
where $\mathbf{H_\text{true}} $ and $\mathbf{H_\text{est}}$ are the true and estimated channel matrices, respectively. 

 We consider an operating frequency is $28$ GHz,  corresponding to a wavelength of $\lambda = 10.71 $ mm. We set the number of TX and RX arrays to 16, with a distance of $10\lambda$ between them, and antenna spacing in the transmit and the receive arrays is $d=\frac{\lambda}{2}$. The total number of downlink and uplink UEs is set to $K=8$ 
{and we assume all UEs are uniformly distributed within 20 m$^{2}$}.
 The path loss parameters are based on experimental results from \cite{7504435}, where the path loss constant at the reference distance is $\Gamma = -72 $ dB, the path loss exponent is $\eta = 2.92$, and the shadow fading standard deviation is $\sigma_{\text{sf}} = 8.7$ dB. Unless otherwise claimed, the number of multi-path components is $P=5$ and AS is $\theta_{\text{AS}}=60^{\circ} $. The SI Rician factor is $\kappa = 40$ dB and we assume that the propagation SI cancellation is $\epsilon_{\text{SI}} = -40$ dB {\cite{9139277}}. 

The NN-based channel estimators are trained on Python $3.7.16$ and implemented using the Keras libraries with a TensorFlow backend in the Jupyter Notebook environment. We employ Adam optimizer with a batch size of $512$  to update the network parameters. A dataset of $50,000$ samples is collected based on the channel model and it is split into $20,000$ samples for training, $20,000$ samples for validation, and $10,000$ samples for testing. The validation data is used to ensure that the model does not simply memorize the training data but learns meaningful aspects of the data for effective prediction.

\subsection{Analysing NN-based channel estimator}
Before examining the different channel estimators for various pilot dimensions discussed in Section \ref{Pilot_transmission_schemes}, we conducted several experiments to understand the behavior of NN-based channel estimators under different channel conditions and NN architectures for training. The following results are specific to SI channel estimation with a pilot dimension of $\tau = N_\text{t} + K$. For the sake of brevity, we have not included plots for other pilot dimensions and UE channel estimation. For the MMSE channel estimator, we obtain the empirical channel covariance matrix estimated through $M$ channel realizations, i.e.,
\begin{equation}
    \mathbf{R}_{q} = \frac{1}{M}\sum_{i=1}^{M} vec(\mathbf{H}_{q}^{(i)})vec(\mathbf{H}_{q}^{(i)})^{H},
\end{equation}
where throughout our simulation, we set $M=1000$.

\begin{figure}
    \centering
    \includegraphics[height=6.5cm,width=8.89cm]{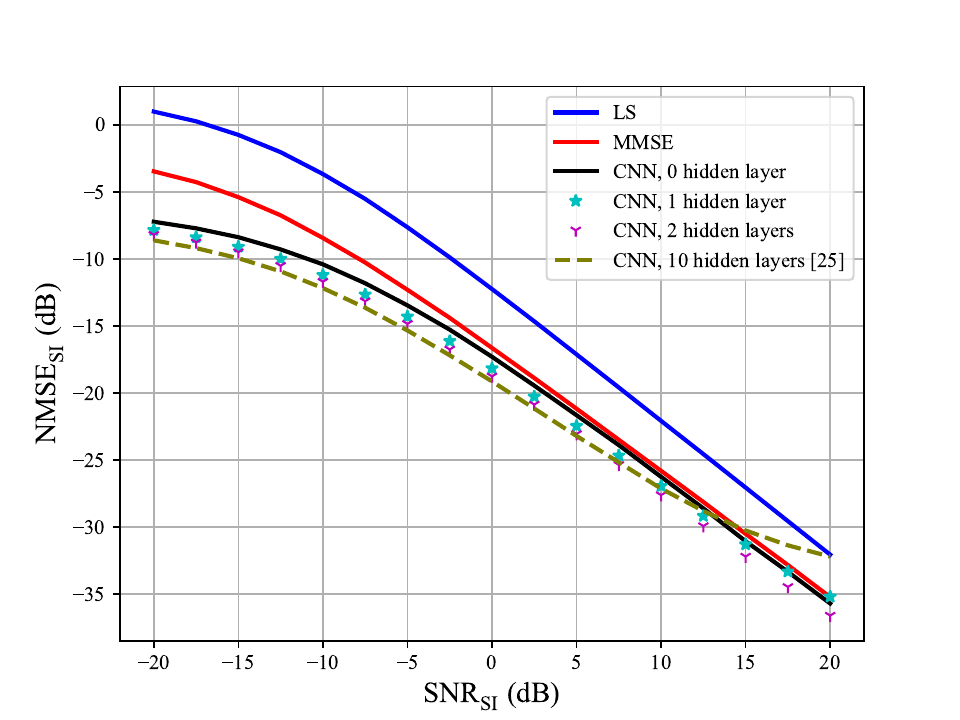}
    \caption{{NMSE\textsubscript{SI} vs SNR\textsubscript{SI} with a varying number of hidden layers.}}
    \label{fig: different_NN_architecture}
\end{figure}

\begin{figure}
    \centering
    \includegraphics[height=6.5cm,width=8.89cm]{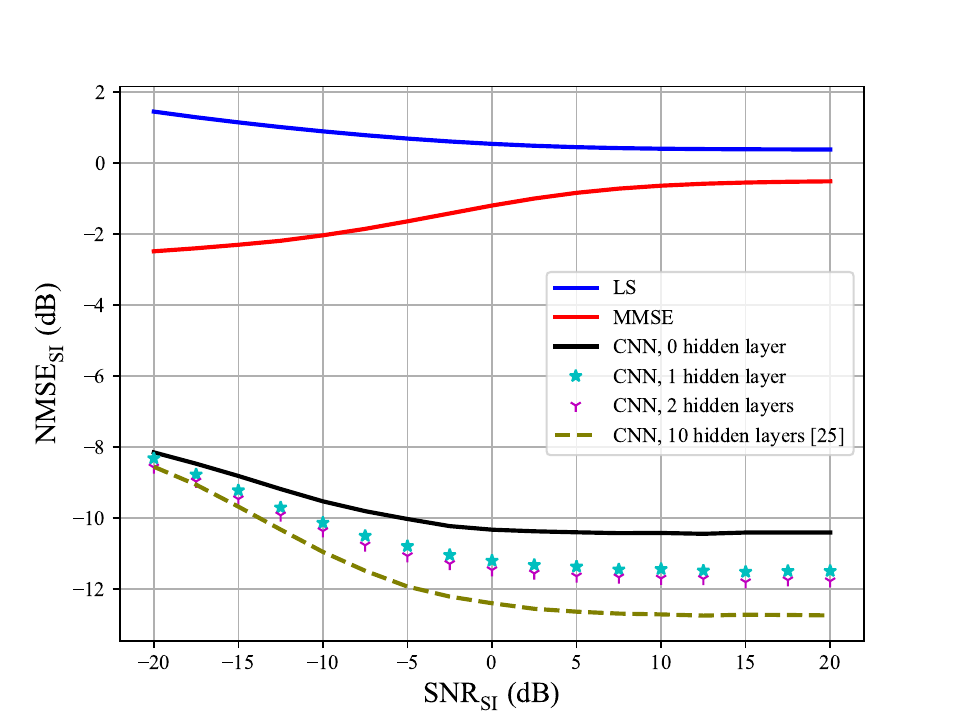}
    \caption{{NMSE\textsubscript{SI} vs SNR\textsubscript{SI} with a varying number of hidden layers for a full-duplex BS with 1-bit ADCs.}}
    \label{fig: 1-bit ADCs}
\end{figure}

Fig. \ref{fig: different_NN_architecture} illustrates the NMSE\textsubscript{SI} versus SNR\textsubscript{SI} with different numbers of hidden layers. We can observe that NN-based channel estimator outperforms the LS and MMSE estimation. Furthermore, utilizing hidden layers slightly improves the NMSE of channel estimates. For instance, CNN with 2 hidden layers decreases the NMSE by about 1 dB compared to CNN with no hidden layer. However, comparing CNN with 10 hidden layers (DeepCNN) with that of 1 or 2 hidden layers suggests that using very deep CNNs does not noticeably improve the channel estimates (about 0.3 dB NMSE improvement), while significantly increasing computational complexity. 
Furthermore, the erratic behavior of DeepCNN at high SNRs can be rectified through fine-tuning, for example, by using a different number of convolutional kernels, adjusting the learning rate, etc. However, for the sake of comparison, we applied the exact same parameters to all CNNs with different numbers of hidden layers.

In practical scenarios, various non-linear distortions in hardware components exist, adding complexity to the problem of channel estimation. To further analyze the impact of the number of hidden layers, we consider the introduction of 1-bit ADCs at the BS to incorporate a non-linear distortion effect into the received pilot signal. When employing 1-bit ADCs, the received pilot signal becomes \cite{9067011}
\begin{equation}\label{1-bit ADC RX signal}
\mathbf{Y_\text{1-bit}} = \text{sgn}(\mathbf{Y_\text{SI}}),
\end{equation}
where $\text{sgn}(\cdot)$ is the element-wise signum function.

Once again, we apply CNN with a varying number of hidden layers to address the channel estimation problem with 1-bit ADCs. In this case, the input to the CNN is $\mathbf{\hat{Y}_\text{1-bit}} = \mathbf{Y_\text{1-bit}}\mathbf{W}^H_{N_\text{t}\times \tau}$, and the output is the estimated channel. Fig. \ref{fig: 1-bit ADCs} presents the NMSE of LS, MMSE, and NN-based channel estimators.
As observed in the figure, the addition of hidden layers improves the NMSE by approximately 1 dB, which is similar to the improvement observed in the case of infinite-bit ADCs in Fig. \ref{fig: different_NN_architecture}.

{

To delve deeper into the problem of channel estimation with low-resolution ADCs, we followed the study in \cite{9847603}. We assume that ADCs apply \( b \)-bit uniform scalar quantization with a set of \( 2^{b} - 1 \) thresholds denoted as \( \{\tau_{1}, \ldots, \tau_{2^{b}-1}\} \). The thresholds of the uniform quantizer will be
\begin{equation}
    \tau_{l} = (-2^{b-1} + l) \Delta_b , \;\;\; l \in L = \{1, \ldots, 2^{b} - 1\},
\end{equation}
where $\Delta_b$ is the step size.

The quantization function is independently applied to the real and imaginary components of the signal, with its definition as follows:
\begin{equation}
    Q_{b}(r) = \begin{cases} 
\tau_{l} - \frac{\Delta_b}{2} & \text{if } r \in (\tau_{l-1}, \tau_{l}] \text{ with } l \in L \\
(2^{b} - 1) \frac{\Delta_b}{2} & \text{if } r \in (\tau_{2^{b}-1}, \tau_{2^{b}}]
\end{cases}
\end{equation}

Therefore, the received pilot signal after b-bit ADC follows:
\begin{equation}\label{n-bit ADC RX signal}
\mathbf{Y_\text{b-bit}} =Q_{b}(\mathbf{Y_\text{SI}}),
\end{equation}

The quantization step size is usually chosen to minimize the quantization MSE, and the optimal values are listed in Table \ref{tab:step_size} for their corresponding bits. The quantization distortion, $\eta_b$ is defined as
\begin{equation}
    \eta_b \triangleq \mathbb{E}\left[ \frac{||Q_b(r) - r||^2_F}{||r||^2_F}\right],
\end{equation}

\begin{table}
\centering
\caption{Optimal uniform quantizer \cite{9847603}}
\label{tab:step_size}
\begin{tabular}{@{}cccc@{}}
\toprule
Resolution (b) & Step size ($\Delta_b$) & Distortion ($\eta_b$) \\
\midrule
1-bit & $\sqrt{\frac{8}{\pi}}$ & $1 - \frac{2}{\pi}$ \\
2-bit & 0.996 & 0.1188 \\
3-bit & 0.586 & 0.0374 \\
4-bit & 0.335 & 0.0115 \\
\bottomrule
\end{tabular}
\end{table}
\begin{figure}
    \centering
    \includegraphics[height=6.5cm,width=8.89cm]{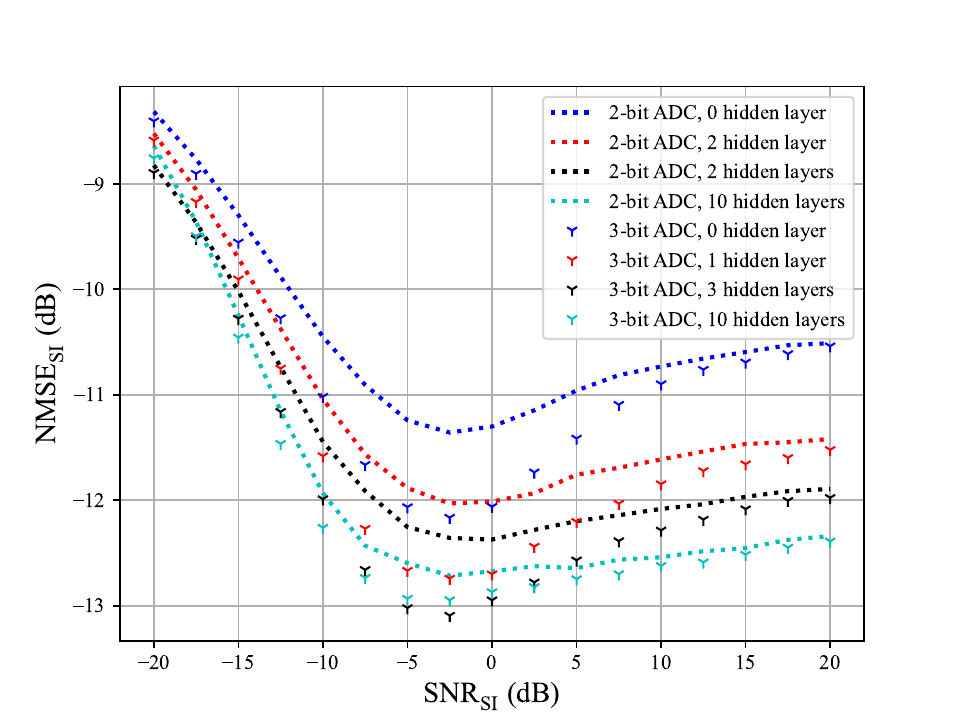}
    
    \caption{{NMSE\textsubscript{SI} vs SNR\textsubscript{SI} with a varying number of hidden layers for a full-duplex BS with few-bit ADCs.}}
    \label{fig:nbit_ADC}
\end{figure}

We conducted simulations for 2-bit and 3-bit ADCs, and similar results can be extrapolated for higher resolutions. From Fig. \ref{fig:nbit_ADC}, we can observe that increasing the number of hidden layers improves estimation with 2-bit and 3-bit ADCs, similar to the results for the 1-bit case. Furthermore, as expected, the higher the resolution of ADCs, the better the NMSE of channel estimates. 
}

\begin{figure*}
    \centering
  \subfloat[$\theta_{\text{AS}}=360^{\circ} $  \label{fighigh_corr_a}]{%
       \includegraphics[height=6.5cm,width=8.89cm]{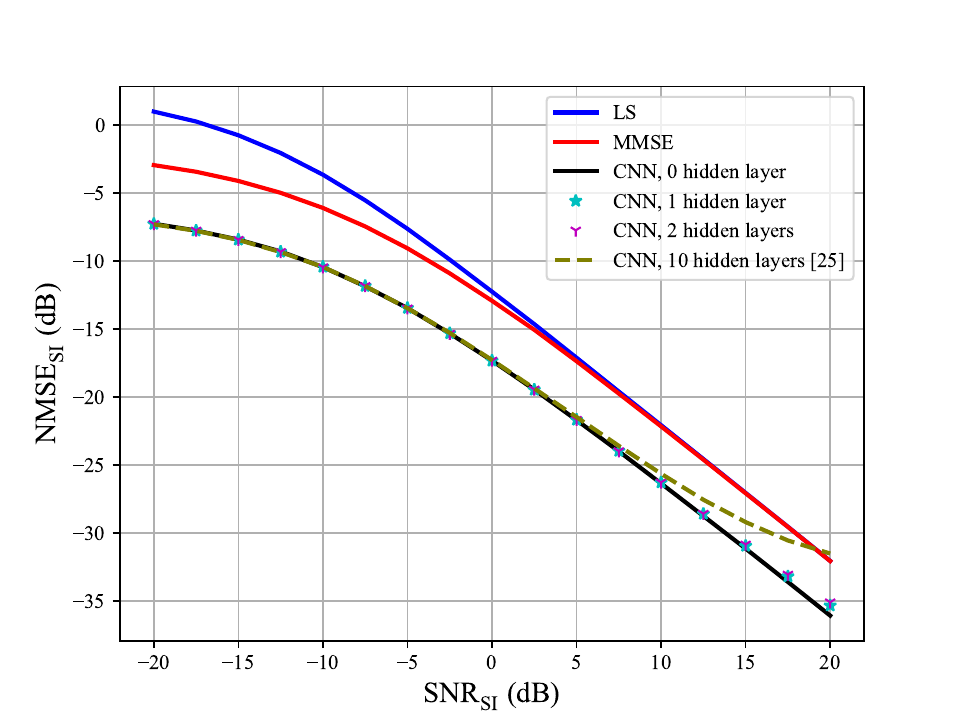}}
  \subfloat[$\theta_{\text{AS}}=10^{\circ} $ \label{high_corr_b}]{%
        \includegraphics[height=6.5cm,width=8.89cm]{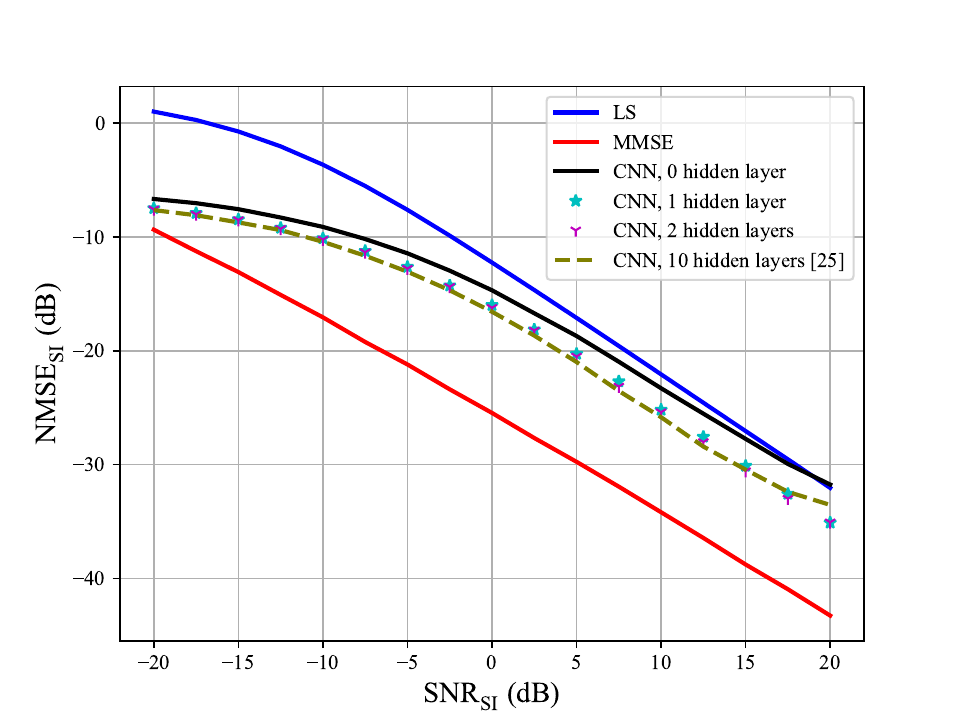}}

  \caption{{NMSE\textsubscript{SI} vs SNR\textsubscript{SI} for (a) low-correlated and (b) high-correlated channels.}}
  \label{fig: uncorrelated and correlated} 
\end{figure*}
Then, we examine the behavior of the NN-based channel estimator under two distinct channel correlation conditions: high-correlated and low-correlated scenarios. The plots in Fig. \ref{fig: uncorrelated and correlated} illustrate how the LS, MMSE, and NN-based estimators perform under different spatial channel correlation strengths. Larger $\theta_{\text{AS}}$ corresponds to lower spatial correlations, and vice versa. In the high-correlated scenario, the MMSE estimator can explicitly leverage the channel covariance matrix, resulting in a significant improvement in estimation quality. This leads to a substantial gap between the LS and MMSE estimations. However, in the low-correlated channel, both the LS and MMSE estimations converge to the same NMSE at high SNRs. Comparing the NN-based estimation with the LS and MMSE estimations in both low-correlated and high-correlated channel conditions provides interesting insights into the behavior of the NN-based estimator. In the low-correlated scenario, the NN-based estimator outperforms both the MMSE and LS estimations. However, in highly correlated channel conditions, the MMSE estimator consistently outperforms the NN-based estimator. This observation suggests that the NN-based estimator, in a data-driven fashion, struggles to utilize the second-order statistics of the channel for estimation, as well as the MMSE estimator does.
Furthermore, the plots suggest that in a low-correlated scenario, increasing the number of hidden layers does not yield any noticeable improvement. However, in a highly correlated channel condition, the NMSE decreases with the addition of more hidden layers to the NN architecture. 
{

Therefore, the choice of the depth of NN architecture is subject to the channel condition and cannot be regarded as universally applicable. Nevertheless, this observation underscores the significance of channel statistics in determining the architecture of NN for the channel estimation problem.
}

{

Finally, Fig. \ref{fig: kappa} shows the NMSE of SI channel estimation versus $\kappa$ with different values of $\kappa$ during training ($\kappa_{train}$). We exclude the plots with 1, 2, and 10 hidden layers for enhanced clarity. As it can be seen from this figure, depending on the value of $\kappa_{train}$, the NMSE of SI channel estimates behaves differently. More specifically, when it is trained on low $\kappa_{train}$, the NMSE\textsubscript{SI} increases as $\kappa$ increases and when it is trained on high $\kappa_{train}$, the NMSE\textsubscript{SI} decreases as $\kappa$ increases. On the other hand, training on different values of $\kappa_{train}$ ranging from 20 to 100 dB, shows that higher values of $\kappa$ lead to higher NMSE\textsubscript{SI}.
\begin{figure}
    \centering
    \includegraphics[height=6.5cm,width=8.89cm]{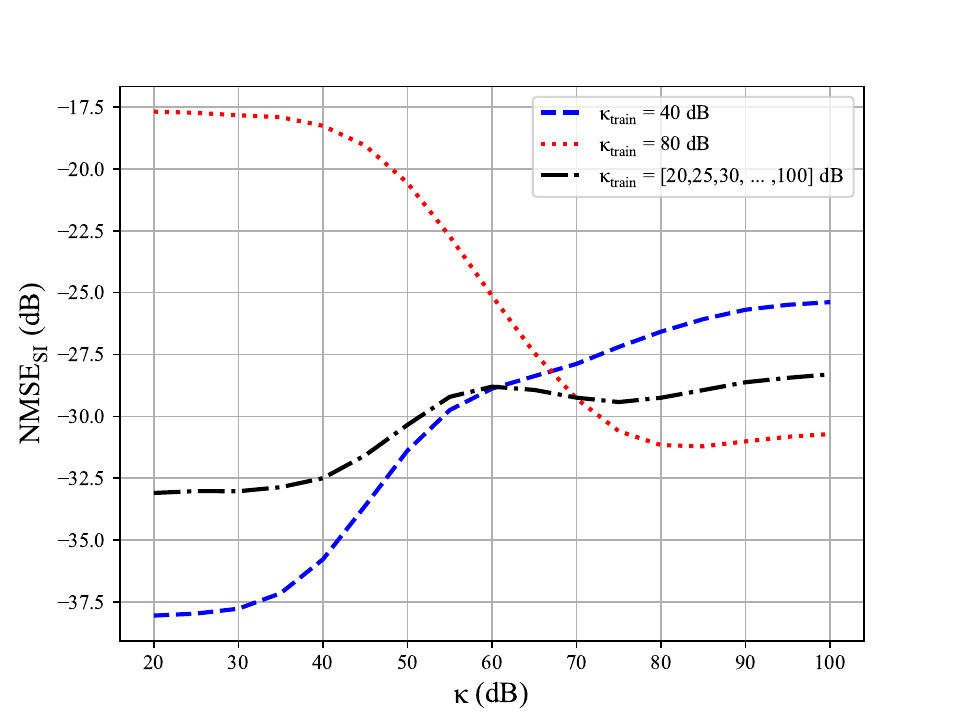}
    \caption{{NMSE\textsubscript{SI} vs $\kappa$ with different values of $\kappa$ during training ($\kappa_{train}$).}}
    \label{fig: kappa}
\end{figure}
}

\subsection{SI and UE channel estimation for different pilot dimensions}
\begin{figure}
    \centering
    \includegraphics[height=6.5cm,width=8.89cm]{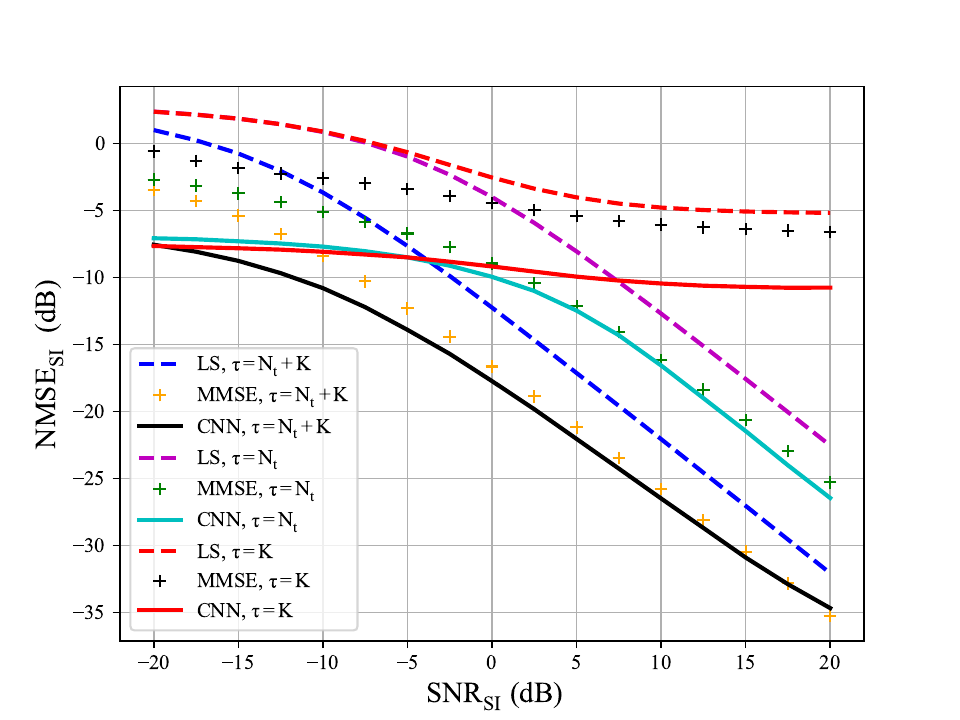}
    
    \caption{NMSE\textsubscript{SI} vs SNR\textsubscript{SI} for different pilot dimensions, SNR\textsubscript{UE} = 0 dB.}
    \label{fig: NMSE_SI_vs_SNR_SI}
\end{figure}

\begin{figure}
    \centering
    \includegraphics[height=6.5cm,width=8.89cm]{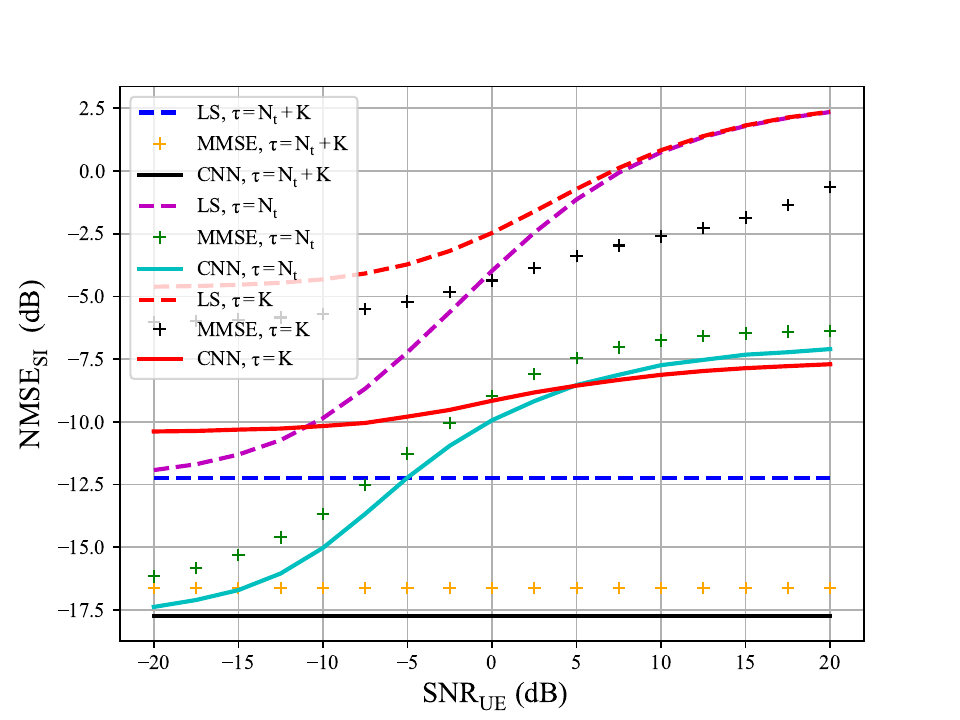}
    
    \caption{NMSE\textsubscript{SI} vs SNR\textsubscript{UE} for different pilot dimensions, SNR\textsubscript{SI} = 0 dB.}
    \label{fig: NMSE_SI_vs_SNR_UE}
\end{figure}

Next, we examine the performance of SI channel estimation for different pilot dimensions, as described in Section \ref{Pilot_transmission_schemes}. For the following simulations, we applied a CNN with no hidden layers and set $\theta_{\text{AS}}=60^{\circ}$.

The NMSE of SI channel estimation vs SNR\textsubscript{SI} and SNR\textsubscript{UE} are provided in Figs. \ref{fig: NMSE_SI_vs_SNR_SI} and \ref{fig: NMSE_SI_vs_SNR_UE} for different pilot dimensions, respectively.
The results demonstrate that the NN-based estimator outperforms the LS and MMSE estimators across considered pilot dimensions. Moreover, from Fig. \ref{fig: NMSE_SI_vs_SNR_UE}, it is evident that the NN-based approach is more resilient to interference from UEs when compared to the MMSE and LS estimation methods.
Using fewer pilot dimensions leads to a significant reduction in the quality of channel estimates for all three approaches. However, the reduction in pilot dimensions can allow more resources to be allocated to transmitting payload data. The trade-off between the performance of channel estimates and the number of pilot dimensions used depends on the required accuracy threshold for channel estimates and the system data rate.

\begin{figure}
    \centering
    \includegraphics[height=6.5cm,width=8.89cm]{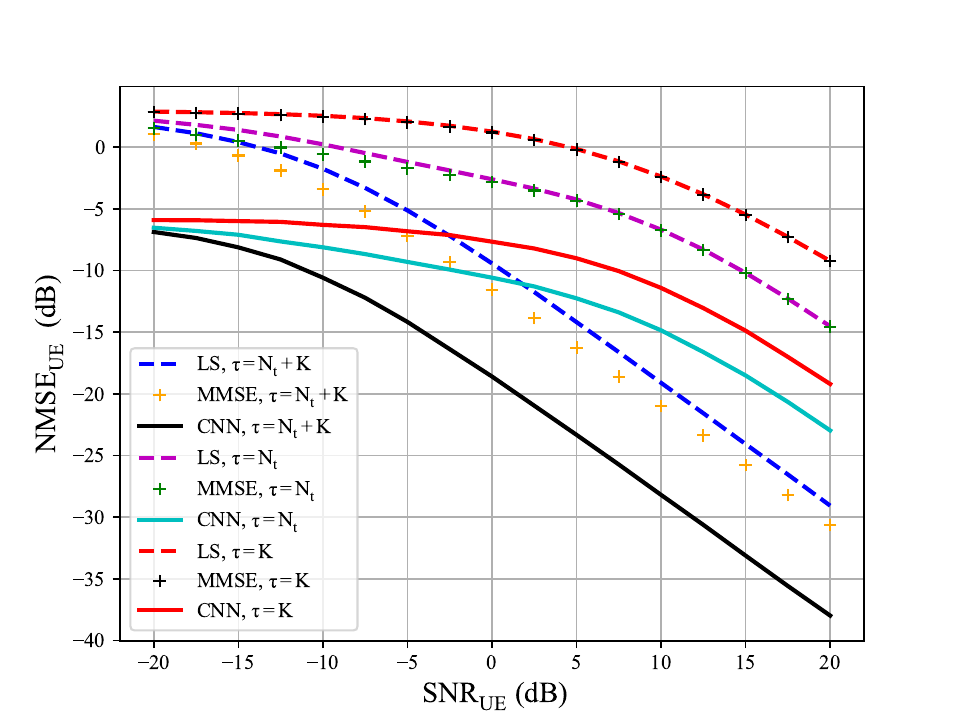}
    \caption{NMSE\textsubscript{UE} vs SNR\textsubscript{UE} for different pilot dimensions, SNR\textsubscript{SI} = 10  dB.}
    \label{fig: NMSE_UE_vs_SNR_UE}
\end{figure}

\begin{figure}
    \centering
    \includegraphics[height=6.5cm,width=8.89cm]{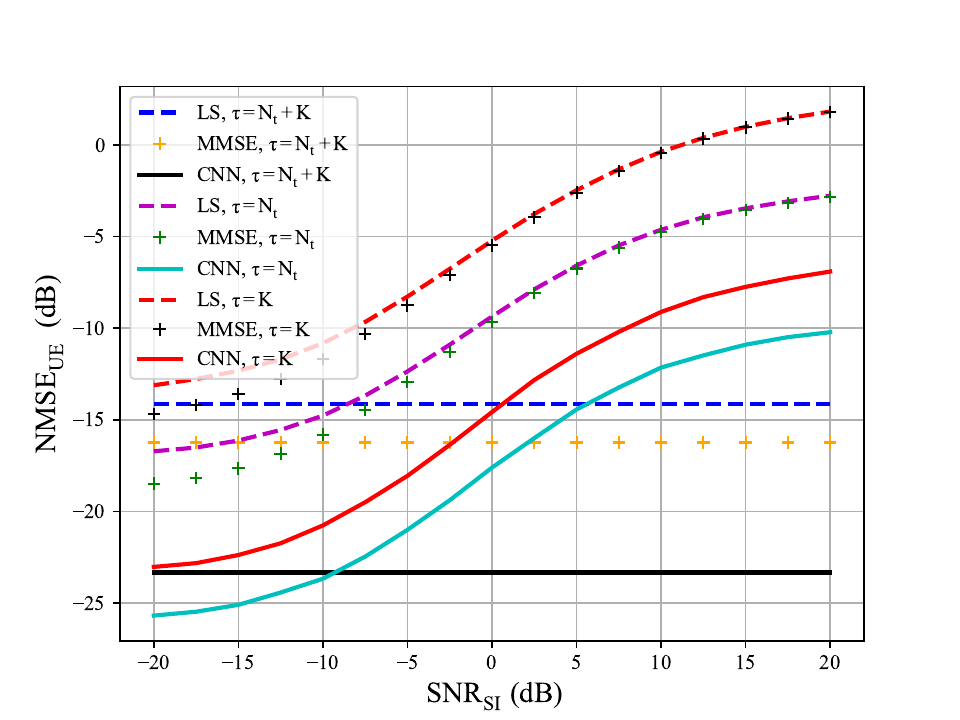}
    \caption{NMSE\textsubscript{UE} vs SNR\textsubscript{SI} for different pilot dimensions, SNR\textsubscript{UE} = 5 dB.}
    \label{fig: NMSE_UE_vs_SNR_SI}
\end{figure}

We have also generated similar plots for UE channel estimation, showing the NMSE with respect to SNR\textsubscript{UE} and SNR\textsubscript{SI}. The plots are presented in Figs. \ref{fig: NMSE_UE_vs_SNR_UE} and \ref{fig: NMSE_UE_vs_SNR_SI}, respectively, with varying pilot dimensions. We employ the MMSE estimator to cancel out the SI signal from the received pilot signal in LS, MMSE, and NN-based channel estimators for UE channel estimation, specifically when the pilot dimensions are $\tau=N_\text{t}$ and $\tau=K$. Note that we also utilize the error covariance matrix for the MMSE estimator when estimating UE channels as denoted in \eqref{MMSE_UE}.
{From the results, it is evident that the NN-based approach using shorter pilot dimensions like $\tau = N_\text{t}$ or $\tau=K$, outperforms LS channel estimation with a pilot dimension of $\tau=N_\text{t}+K$. Even the MMSE channel estimator with longer pilot dimensions (e.g., $\tau=N_\text{t}+K$), is unable to outperform the NN-based technique with shorter pilot dimensions (e.g., $\tau=N_\text{t}$). These findings clearly demonstrate the superiority of NN in accurately estimating wireless channels, particularly in scenarios with lower SNRs and higher interference.}

\begin{figure*} 
    \centering
  \subfloat[LS  \label{NMSE_UE_vs_SNR_SI_SNR_UE_CNN_case3_a}]{%
       \includegraphics[width=0.33\linewidth]{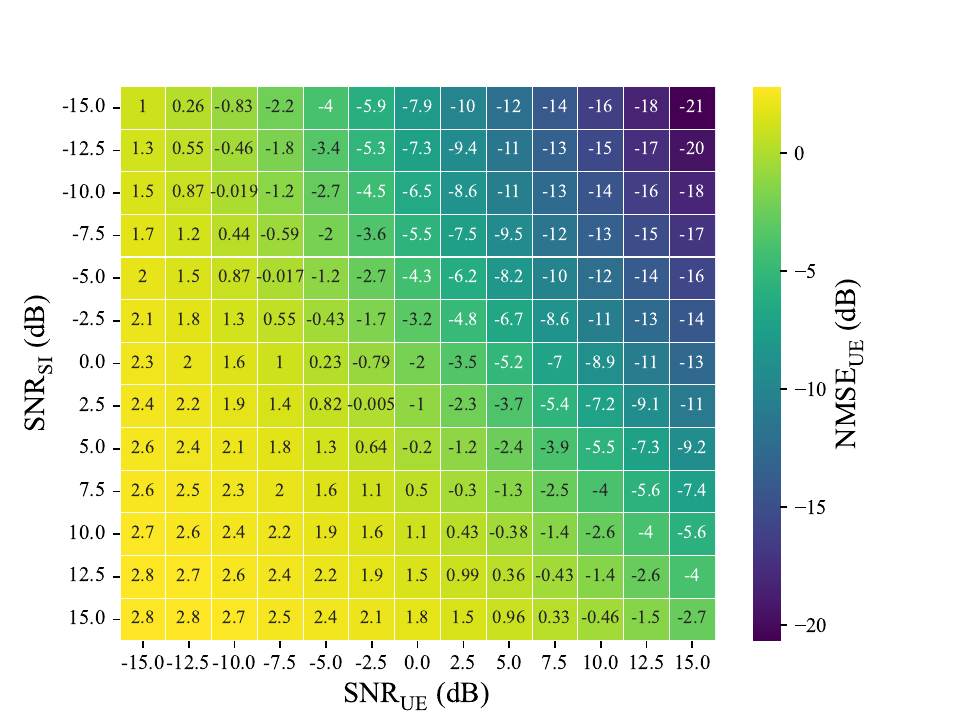}}
    \hfill
  \subfloat[MMSE \label{NMSE_UE_vs_SNR_SI_SNR_UE_CNN_case3_b}]{%
        \includegraphics[width=0.33\linewidth]{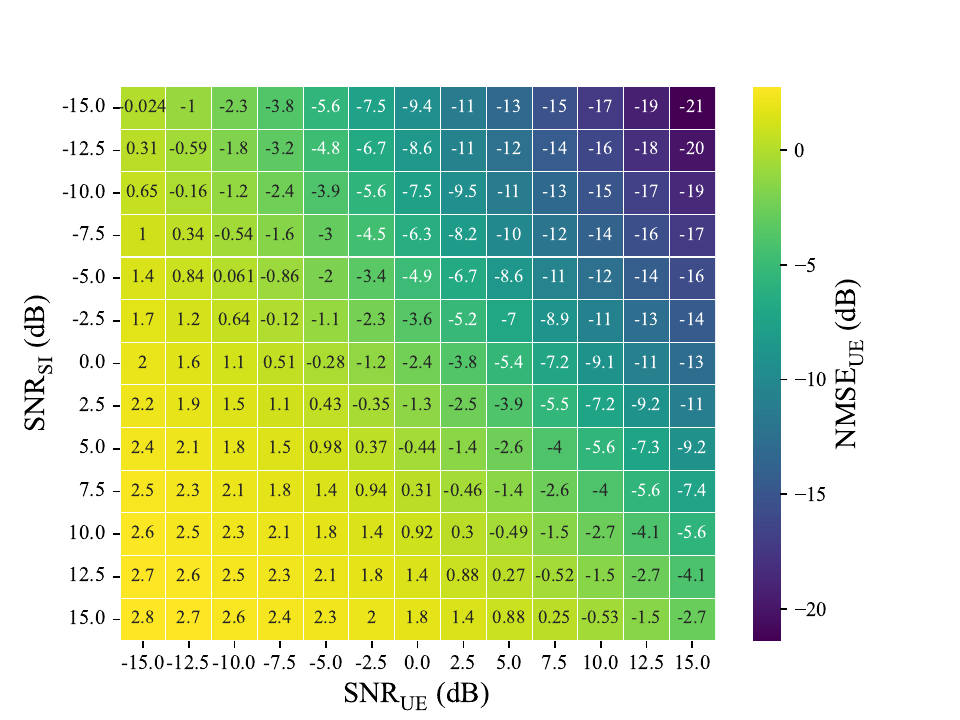}}
    \hfill
  \subfloat[NN \label{NMSE_UE_vs_SNR_SI_SNR_UE_CNN_case3_c}]{%
        \includegraphics[width=0.33\linewidth]{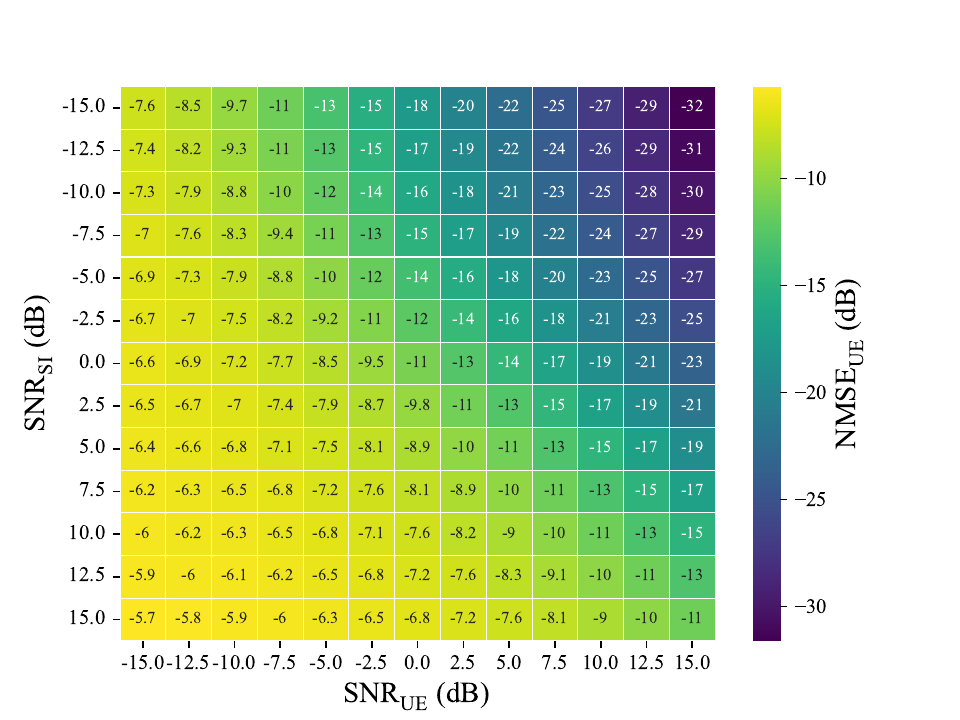}}

  \caption{NMSE\textsubscript{UE} vs SNR\textsubscript{SI} and SNR\textsubscript{UE} for  (a) LS and (b) MMSE (c) NN channel estimators, $\tau = K$.}
  \label{fig: NMSE_UE_vs_SNR_SI_SNR_UE} 
\end{figure*}

A higher SNR\textsubscript{SI} results in better SI channel estimation and, consequently, less error in SI cancellation for UE channel estimation. Conversely, a higher SNR\textsubscript{UE} increases the NMSE of SI channel estimation, leading to an increase in error for UE channel estimation. On the other hand, a higher SNR\textsubscript{UE} leads to improved UE channel estimates, while a higher SNR\textsubscript{SI} increases the power of interference during UE channel estimation. 
To understand the joint effect of SNR\textsubscript{UE} and SNR\textsubscript{SI} for UE channel estimation, we generated the color bar plots in Fig. \ref{fig: NMSE_UE_vs_SNR_SI_SNR_UE} for LS, MMSE, and NN-based estimation considering a pilot dimension $\tau=K$. Based on the results in this figure, we can conclude that a lower SI signal power or higher SI cancelation will lead to better UE channel estimates.

\begin{figure}
    \centering
    \includegraphics[height=6.5cm,width=8.89cm]{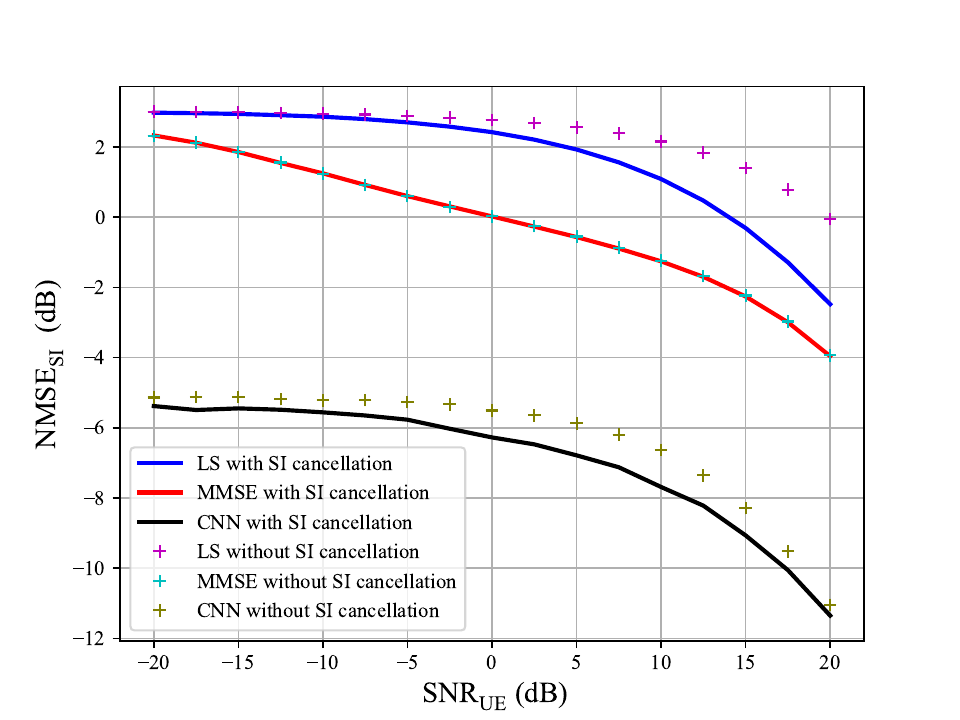}
    \caption{UE channel estimation with/without SI cancellation, $\tau=K$,  SNR\textsubscript{SI} = 20 dB.}
    \label{fig: with/without SI cancelation}
\end{figure}

To further analyze the effect of SI cancelation on UE channel estimation, we considered two cases: in the first case, we cancel out the effect of the SI signal from the received pilot signal with the estimated SI channel, while in the second scenario, we estimate UE channels in the presence of interference from the SI channel. The results are shown in Fig \ref{fig: with/without SI cancelation}. Note that for the MMSE estimation, in the case of SI cancellation, we incoporate the error covariance matrix of the estimated SI channel together with the covariance matrix of the UE channels, while for the scenario without SI cancellation, we exploit the covariance matrix of both SI and UE channels. We can observe that MMSE estimations for these two scenarios result in the same NMSE of UE channel estimates. By comparing the results for LS, MMSE, and NN channel estimators, we observe that SI cancellation during the pilot transmission phase does not provide significant improvement, while estimating the large SI channel matrix is costly.

\begin{figure}
    \centering
    \includegraphics[height=6.5cm,width=8.89cm]{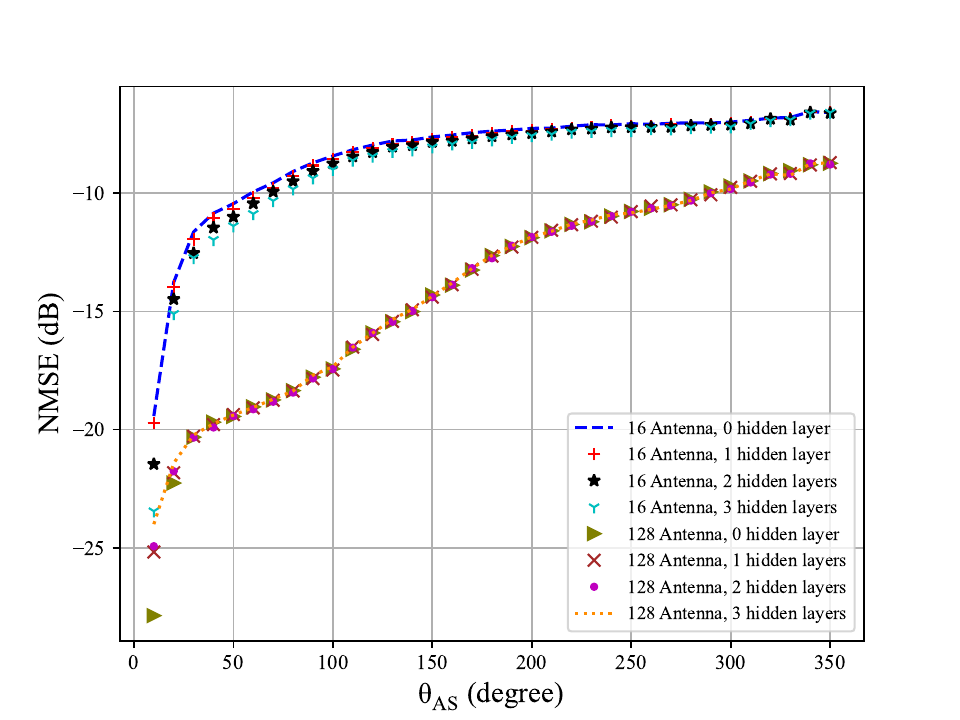}
    \caption{NMSE of channel mapping between RX and TX arrays vs angular spread.}
    \label{fig: RX_TX_Mapping}
\end{figure}
To assess the performance of the NN-based channel mapping between RX and TX arrays at the BS in the separate antenna configuration, we present the NMSE of this mapping concerning varying angular spreads and two different numbers of antennas in Fig. \ref{fig: RX_TX_Mapping}. We trained an FNN with a varying number of hidden layers at angular spreads of $[10, 100, 190, 280]$ degrees during training.
From this figure, we can conclude that an NN effectively predicts the channel from the TX arrays to downlink UEs, given the CSI from downlink UEs to the RX arrays. The prediction accuracy is higher in scenarios with lower angular spread, attributed to increased spatial correlation, and vice versa. Furthermore, the quality of prediction improves with an increased number of antennas, as angular resolutions increase with a larger array. 

{

\subsection{Distribution shift}
\begin{table}[t]
    \centering
    \caption{DeepMIMO dataset parameters}
    \begin{tabular}{ll}
        \toprule
        \textbf{DeepMIMO parameters} & \textbf{Value} \\
        \midrule
        Number of paths & 10 \\
        Active UEs & from row 550 to 1650 \\
        Active BS number & BS 3 \\
        Bandwidth & 50 MHz \\
        BS antennas & $N_x$ = 1, $N_y$ = 16, $N_z$ = 1 \\
        UE antennas & $N_x$ = 1, $N_y$ = 1, $N_z$ = 1 \\
        \bottomrule
    \end{tabular}
    \label{tab:deepmimo}
\end{table}

When a learning model is trained on a particular dataset with a specific distribution of features, it learns to make predictions based on the patterns inherent in that dataset. However, deploying such models in real-world scenarios often involves testing them with datasets that may differ in distribution from the training data. This misalignment between training and test distributions can lead to a phenomenon known as distribution shift, where the model's performance deteriorates due to its inability to generalize effectively to unseen data and it poses significant challenges in testing the efficacy and reliability of learning models.

To analyze the effect of distribution shift on our trained models, we utilize the DeepMIMO dataset in the test phase. This dataset constructs the MIMO channels based on ray-tracing simulation from Remcom Wireless InSite \cite{Remcom}. Specifically, we use the ‘O1’ scenario at 28 GHz with the parameters specified in Table \ref{tab:deepmimo}. We generate 4977 channel realizations from the DeepMIMO dataset to test the NN model for UE channel estimation trained based on the channel data from geometrical channel models introduced in Section \ref{section2_channel_modeling}. The same can be applied for SI channel estimation, but for brevity, we do not include the results for this case. As seen from Fig. \ref{fig: Distribution_shif}, due to distribution mismatch, the NMSE degrades in the test phase. Furthermore, we observe that NNs with fewer hidden layers generalize better compared to deeper CNNs since the number of trainable parameters is much higher in deep CNNs. This underscores the benefits of utilizing simpler NN architectures in terms of generalization abilities.

\begin{figure}
    \centering
    \includegraphics[height=6.5cm,width=8.89cm]{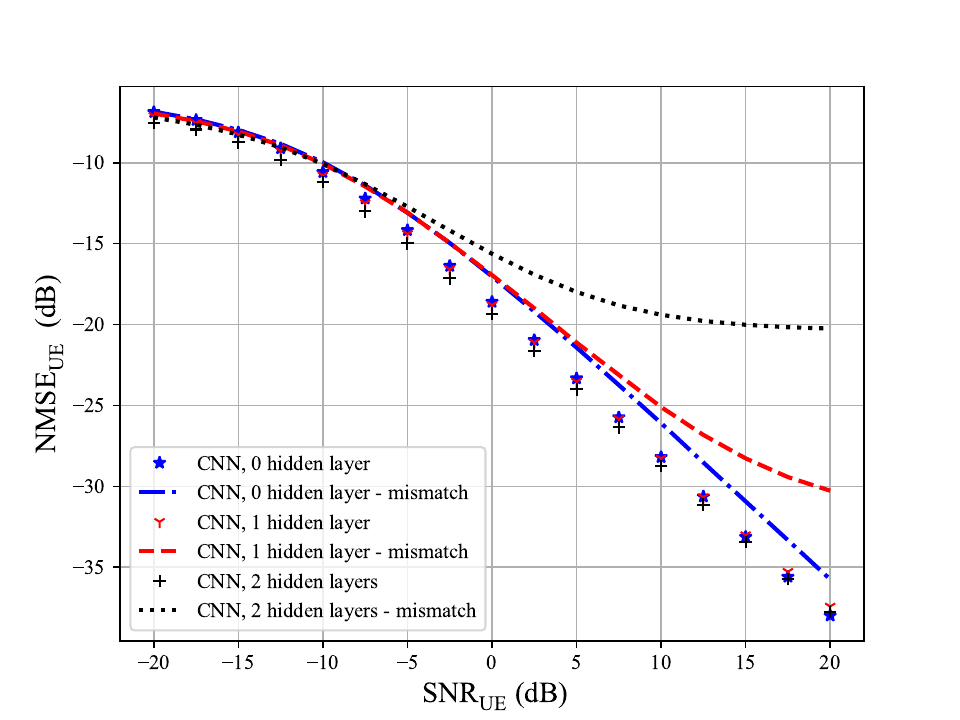}
    
    \caption{Performance of NN channel estimator for UE channel estimation with distribution shift.}
    \label{fig: Distribution_shif}
\end{figure}
}
\subsection{Complexity Analysis}
\begin{figure}
    \centering
    \includegraphics[height=6.5cm,width=8.89cm]{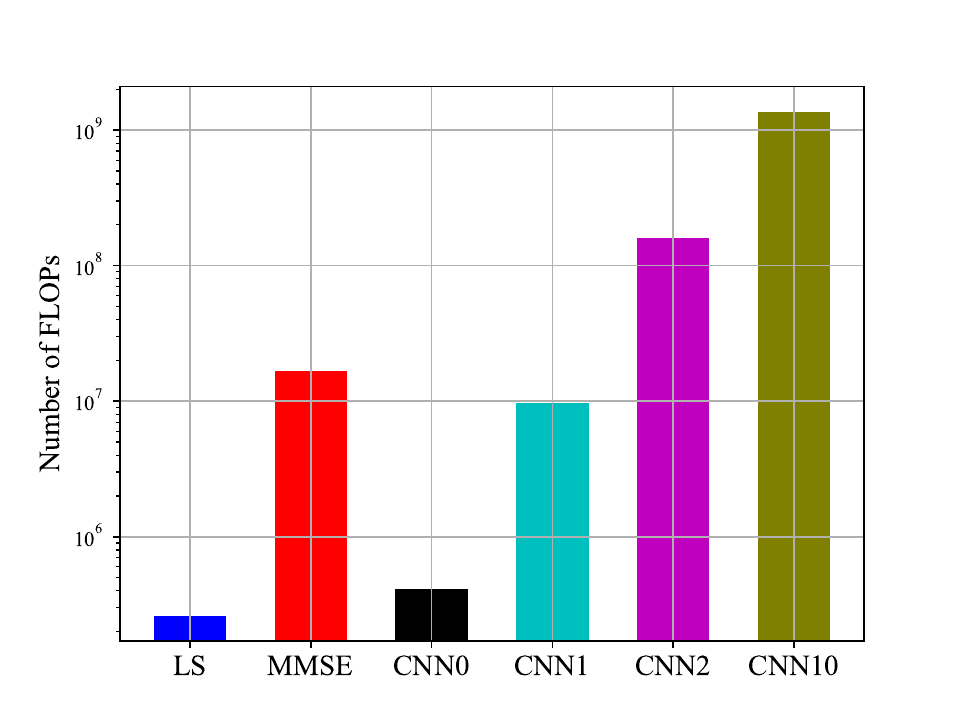}
    
    \caption{Number of FLOPs for SI channel estimation of different estimators. CNNx represents a CNN with x hidden layer(s).}
    \label{fig: Complexity_Analysis}
\end{figure}

Finally, we compare the computational complexity of the LS, MMSE, and NN channel estimators in terms of FLOPs. The results are summarized in Table \ref{Complexity_table}, where $\zeta$ represents the side length of the convolutional window, and $f_{l-1}$ and $f_l$ denote the numbers of input and output convolutional kernel of the $l$-th layer. A bar plot is shown in Fig. \ref{fig: Complexity_Analysis} for SI channel estimation and pilot dimension $\tau=N_\text{t}$ based on the CNN architecture used in the simulation, i.e., $\zeta = 3$, $f_0=f_{L} = 2$, and $f_i= 64 $ for the remaining layers ($i=1,2,\dots, L-1$). Note that we have ignored the computational complexity of the covariance matrix calculation for MMSE estimation, as the channel statistics do not typically change during several coherence blocks. For the same reasons, we also ignored the computational complexity of the matrices in the MMSE formula that depend on spatial correlation matrices, assuming they can be pre-computed and used until the channel statistics change substantially. 
According to table \ref{Complexity_table}, the number of FLOPs exhibits quadratic growth with the number of received antennas in MMSE estimation, while for LS and NN-based estimation, it is linear. Furthermore, CNNs with no or one hidden layer require fewer FLOPs than MMSE estimation, while increasing the number of hidden layers significantly increases the computational complexity. In particular, adding a hidden layer with $64$ convolutional kernels and a $3\times3$ window size requires about $1.5\times 10^8$ more FLOPs for estimating a MIMO channel with $16$ antennas.

\begin{table}
    \centering
    \caption{Number of FLOPs for  SI and UE channel estimation using  LS, MMSE, and NN estimators.}
    \begin{tabular}{ |c|c|c|c| }
    \hline
    &LS & MMSE & CNN \\ 
    \hline
    SI  &$ N_\text{r}N_\text{t}  \tau $ & $ N_\text{r}^2N_\text{t}\tau  $ & $ N_\text{r}N_\text{t}  \tau + N_\text{r} N_\text{t} \sum_{l=1}^{L} \zeta^2 f_{l-1}f_l$  \\ 
    \hline
    UE  &$N_\text{r}K  \tau $ & $N_\text{r}^2K\tau $ & $ N_\text{r}K  \tau + N_\text{r} K \sum_{l=1}^{L} \zeta^2 f_{l-1}f_l$ \\ 
    \hline
    \end{tabular}
    
    \label{Complexity_table}
\end{table}

\section{Conclusion}\label{section5}
{

In this paper, we studied the channel estimation problem for full-duplex mmWave MIMO systems using NNs. We show that simple NNs with no or few hidden layers can achieve comparable NMSE to deep NNs. This conclusion holds to some extent even in the presence of few-bit ADC distortion, with additional hidden layers only slightly improving the NMSE of channel estimates. Furthermore, simpler NN architectures with fewer hidden layers demonstrate more robust generalization abilities compared to deep NNs. The comparison between the NN and MMSE channel estimators indicates a notable difference. In the high-correlated regime, MMSE is the preferable channel estimator compared to LS and NN, as it can leverage the explicit channel covariance matrix.
However, in scenarios where spatial correlation is not extremely high, our results demonstrate that NNs outperform MMSE estimation. Additionally, the NN-based channel estimator performs remarkably better than MMSE when fewer pilot resources are utilized for channel estimation. This performance allows for a reduction in pilot overhead in full-duplex systems. We applied FNNs to approximate RX-TX channel mapping for the separate antenna configurations in full-duplex systems. Simulation results demonstrate that NNs effectively map channels from downlink UEs to the receive arrays to the channel from the transmit arrays to downlink UEs, particularly in scenarios with high correlation and large antenna arrays.
Finally, the complexity analysis reveals that NN-based estimation with fewer than two convolutional hidden layers requires fewer FLOPs compared to MMSE estimation.
}

\bibliographystyle{IEEEtran}
\bibliography{Mybib}

\end{document}